%% file: MS.tex
\documentclass[reprint,superscriptaddress,amsmath,amssymb,aps,]{revtex4-2}

\usepackage{graphicx}
\usepackage{amsmath}
\usepackage{amssymb}
\usepackage{dcolumn}
\usepackage{bm}
\usepackage{mathtools}
\usepackage{hyperref}
\usepackage{comment}
\usepackage{xcolor}
\usepackage[utf8]{inputenc}
\DeclareMathOperator{\sign}{sign}

\usepackage{subfiles}

\begin{document}
\UseRawInputEncoding

\title{On-Site Potential Creates Complexity in Systems with Disordered Coupling}

\author{I. Gershenzon}
\affiliation{Department of Physics of Complex Systems, Weizmann Institute of Science, Rehovot 7610001, Israel}

\author{B. Lacroix-A-Chez-Toine}
\affiliation{Department of Mathematics, King’s College London, London
WC2R 2LS, United Kingdom}
\affiliation{Department of Physics of Complex Systems, Weizmann Institute of Science, Rehovot 7610001, Israel}

\author{O. Raz}
\affiliation{Department of Physics of Complex Systems, Weizmann Institute of Science, Rehovot 7610001, Israel}

\author{E. Subag}
\affiliation{Department of Mathematics, Weizmann Institute of Science, Rehovot 76100, Israel}

\author{O. Zeitouni}
\affiliation{Department of Mathematics, Weizmann Institute of Science, Rehovot 76100, Israel}


\begin{abstract}

We calculate the average number of critical points $\overline{\mathcal{N}}$ of the energy landscape of a many-body system with disordered two-body interactions and a weak on-site potential. We find that introducing a weak nonlinear on-site potential dramatically increases $\overline{\mathcal{N}}$ to exponential in system size and give a complete picture of the organization of critical points. Our results extend solvable spin-glass models to physically more realistic models and are of relevance to glassy systems, nonlinear oscillator networks and many-body interacting systems.

\end{abstract}

\maketitle



\paragraph*{Introduction}
The interplay between coupling, disorder and non-linearity is of interest in diverse areas of science. In physics, notable examples occur in spin and structural glasses \cite{Castellani2005Spin-glassPedestrians,mezard1987spin,Diep2013,Gartner2016NonlinearGlasses,Goremychkin2008Spin-glassFrustration,Kim2010}, many-body localisation \cite{Tikhonov2021FromLocalization,Abanin2019Colloquium:Entanglement},  nonlinear wave propagation in disordered media \cite{Antenucci2015ComplexMedia,Zyuzin2004PropagationMedia,Conti2011ComplexityMedia,Ghofraniha2015ExperimentalLasers,Shaw1993NormalSystems,Fishman2012ThePuzzles}, ``dirty" superconductors \cite{Kirkpatrick1992SuppressionDisorder,Fisher1991ThermalSuperconductors}, coupled oscillator networks \cite{Inagaki2016,Wang2013,Nixon2013,Gopalakrishnan2011FrustrationInteractions},  atomic spin gases \cite{Horowicz2021CriticalGas,Lippe2021ExperimentalModel} and in other systems \cite{Fort2005EffectWaveguide,Lugan2007UltracoldCondensate,Tsui1999NobelFields,Joshi2020DeconfinedMagnet,Eilbeck1985TheEquation,Goldstein1969ViscousPicture}. Non-linearity, coupling and disorder are also commonly found in sociological models \cite{Namatame2016Agent-BasedDynamics,Palla2007CommunityNetworks,Ebel2002DynamicsNetworks}, epidemiology \cite{Sattenspiel1995ARegions,Pastor-Satorras2015EpidemicNetworks,Fu2011ImitationNetworks,Roddam2001MathematicalInterpretation,Hethcote2000MathematicsDiseases}, ecological systems \cite{May2001StabilityBiology,Fyodorov2016NonlinearTransition,Allesina2015ThePerspective}, computer-science \cite{Bahri2020StatisticalLearning,Baity-Jesi2019ComparingSystems,Becker2020GeometryNetworks,Mezard2009InformationComputation} and many other systems \cite{Newman2011TheNetworks}. 

The energy of such disordered systems commonly exhibits a ``rugged'' landscape with a number of critical points that scales exponentially with system size. The abundance of critical points of certain energy and index is quantified in this work by the ``complexity" i.e. the exponential scaling coefficient with system size \cite{Bray2007StatisticsSpaces} (sometimes referred to as the configurational entropy). This is defined as 
\begin{equation}\label{eq:Def_Complexity}
    \Sigma_{k}\left(E\right) = \lim_{N\rightarrow\infty}\frac{1}{N} \log\overline{\mathcal{N}_{k}\left(E\right)},
\end{equation}
where $\overline{\mathcal{N}_k\left(E\right)}$ is the disorder averaged density of critical points of index $k$ at energy $E$. The geometry of a system's energy landscape, described by $\Sigma_{k}\left(E\right)$, was shown to directly influence static and dynamic properties of complex systems such as the mechanical properties of amorphous solids \cite{Bouchbinder2021,Jin2017}, the structure, function and thermal properties of bio-molecules \cite{Milanesi2012_Protein_Landscape_Roughness}, pinning properties of polymers to surfaces \cite{Fyodorov2018,Fyodorov2020ManifoldsDepinning}, the heat capacity of bio-molecules \cite{Goldstein1976ViscousLiquid}, the relaxation time scales of glassy systems \cite{Nishikawa2022,Folena2020,Rizzo2021,Cavagna2001RoleTransition,Buchenau2003EnergyGlasses,Buchenau2003EnergyGlasses}, the mobility in glass-forming liquids \cite{Charbonneau2014}, the transition rates between meta-stable states in complex systems \cite{Milanesi2012_Protein_Landscape_Roughness,Ros2019ComplexitySystems,Rizzo2021}. In addition, the structure of rugged landscapes has a profound influence on optimization algorithms such as deep neural network training \cite{Dauphin2014IdentifyingOptimization,Bahri2020StatisticalLearning,Becker2020} and combinatorial optimization \cite{Mezard2009InformationComputation}. 

Early works used replica methods to derive approximate expressions for $\Sigma$ in spin-glass models \cite{Castellani2005Spin-glassPedestrians, Cavagna1998StationaryEnergy, Cavagna2001RoleTransition, mezard1987spin,Rosinberg2009HysteresisVersion,Kurchan1991ReplicaEquations,Monasson1995} and for disordered nonlinear optical systems \cite{Conti2011ComplexityMedia, Antenucci2015ComplexMedia}. Recently, rigorous random matrix methods were applied to count critical points in many models, including spin glasses \cite{Fyodorov2004ComplexityMatrices, Fyodorov2007ReplicaComplexity,Auffinger2013RandomGlasses, Auffinger2013ComplexitySphere, Subag2017TheApproach, McKenna2021Non-invariantComplexity, McKenna2021ComplexityGlasses,Arous2021LandscapeManifold, Subag2021TheApproach, Subag2018FreeGlasses,Ros2019ComplexitySystems}, 
ecological systems \cite{Fyodorov2016NonlinearTransition}, neural networks \cite{Becker2020GeometryNetworks, Choromanska2015TheNetworks,Maillard2020} and others \cite{Fyodorov2020ManifoldsDepinning, Lacroix-A-Chez-Toine2022CountingRates}.

Several works studied the complexity in confined disordered models, where a global confining potential term is added to a disordered energy landscape \cite{Fyodorov2004ComplexityMatrices, Fyodorov2007ReplicaComplexity, Arous2021LandscapeManifold,Fyodorov2014TopologyOptimization,DZ15, Belius2022TrivialityField}. However, in many physical problems one is interested in coupled many body systems with an on-site (nonlinear) potential. Notable systems where the local nonlinearity plays a central role alongside disordered coupling include amorphous solids \cite{Bouchbinder2021,Charbonneau2014,Jin2017}, supefluidity, superconductivity and Bose-Eienstein condensates (all modeled by the Gross-Pitaevskii equation) \cite{Kirkpatrick1992SuppressionDisorder,Fort2005EffectWaveguide,Lugan2007UltracoldCondensate,Shapiro2012}, wave propagation in nonlinear media \cite{Zyuzin2004PropagationMedia,Conti2011ComplexityMedia}, networks of coupled non-harmonic oscillators (optical, electrical, mechanical) \cite{Caravelli2019, Inagaki2016, Nixon2013} and models of conductors \cite{Tarnopolsky2020Metal-insulatorModel}. All these diverse systems share a common model energy structure: the coupling between DOFs is bi-linear while the local (``on-site") potential energy is non-linear.

In this work, we set out to study the complexity in this family of important models by analysing a prototypical model with weak on-site non-linearity (of general form) and disordered bi-linear coupling. 
A central question motivating this study is the following: The bilinear form potential has at most $N$ critical points, thus zero complexity. On the other hand, the limit of strong  on-site potential can give rise to the Ising model with positive complexity \cite{Mezard2009InformationComputation}. It is therefore natural to ask: {at which ratio between the two terms does the energy landscape become complex and what is the nature of this transition?} We answer this question in the regime of weak on-site potential by deriving a perturbative expression of the annealed complexity.

We find that in our model even a weak on-site term brings about positive complexity, with an exponential number of fixed points i.e. a "rugged" energy landscape. We also find that the landscape of the system exhibits a qualitatively different critical point distribution in energy and index in comparison to the unperturbed model. Interestingly, the effect of adding an on-site potential resembles that of adding higher order interaction terms (as in the mixed $p$-spin model). We also find that an external magnetic field leads to a first order phase transition into a {\it trivial phase} with zero complexity (to leading order) and find the critical field. These results imply that disordered systems comprised of coupled nonlinear units are expected to be found in a glassy phase exhibiting aging and memory in its dynamics, non-trivial response to external forces as well as anomalous thermal properties. Technically, the calculations are facilitated by deriving a result on the disorder average of the modulus of the determinant of a sum of a GOE random matrix and a small deterministic diagonal matrix \cite{Arous2021LandscapeManifold}. 

The model analysed in this work is defined by the energy landscape function $H(\sigma)$ \cite{Rosinberg2009HysteresisVersion}: 
\begin{equation}\label{Eq:ModelDef}
    H\left(\sigma\right)  = \sum_{i,j=1}^{N}J_{ij}\sigma_{i}\sigma_{j} + \kappa\sum_{i=1}^{N}u(\sigma_{i})
     \equiv \phi\left(\sigma\right) + \kappa U\left(\sigma\right)
\end{equation}
where $\sigma = \left(\sigma_{1},\cdots,\sigma_N\right)$ is the vector whose components $\sigma_i$ are continuous real valued ``spin" DOFs constrained to the $N$-sphere i.e. $\sum_{i}\sigma_{i}^{2}=N$, $J_{ij}$ is a random coupling matrix modelling disordered coupling, $u(\sigma_i)$ is a deterministic  ``on-site" nonlinear potential (with bounded derivatives), and $\kappa$ is the deterministic potential strength. Note that in this model the number of critical points is invariant to global shifts in $u(x)$ as well as to addition of quadratic terms (which are constant on the $N$-sphere). In this work, we fix the definition of $u(x)$ by assuming that the Gaussian weighted average of $u$ and $u''$ is zero. The stochastic part of $H$ is denoted by $\phi\left(\sigma\right)$ and the deterministic one by $\kappa U\left(\sigma\right)$. We choose the coupling matrix $J$ to be distributed according to the Gaussian orthogonal ensemble \cite{Anderson2009AnMatrices} - a disordered mean-field coupling. This choice along with the spherical constraints allows for the identification of the coupling term in $H$ with the 2-spin spherical model \cite{Auffinger2013RandomGlasses}. 

To evaluate the average total number of critical points $\overline{\mathcal{N}_{\mathrm{tot}}}$ of the model we use the  Kac-Rice formula on the sphere (as given in \cite{Auffinger2013RandomGlasses}):
\begin{equation} \label{eq:Kac_Rice}
    \overline{\mathcal{N}_{\mathrm{tot}}} = \int_{\sqrt{N}\mathbb{S}^{N-1}} d\sigma_{N-1} \overline{\left| \det\left( \nabla^{2}H \right) \right| \delta\left( \nabla H \right)}
\end{equation}
where $\nabla H$ and $\nabla^{2}H$ are the covariant derivative and Hessian matrix of $H$ on the sphere, respectively. In what follows we evaluate the annealed total complexity $\Sigma_{\mathrm{tot}}$ defined similarly to \eqref{eq:Def_Complexity} in the perturbative regime of weak on site potential, i.e. $\kappa \ll 1$. 

\paragraph*{Main Results}
In this section we cite the results for the total annealed complexity $\Sigma_{\mathrm{tot}}$, and the annealed complexity of fixed extensive index $k=Na$ (number of negative eigenvalues of the Hessian) and extensive energy $E = \epsilon N$, $\Sigma_{a}(\epsilon)$ as a functional of the potential $u(x)$. 
The complexity $\Sigma_{\mathrm{tot}}$ is found to be:
\begin{equation} \label{eq:complexity_main_result}
\begin{split}
    & \Sigma_{\mathrm{tot}}[u] = \kappa^{2}\max\left(0,\Theta\left[u\right]\right) + O(\kappa^{3}),
\end{split}
\end{equation}
with
\begin{equation} \label{eq:Theta_def}
    \Theta\left[u\right] = \frac{1}{4}\int dx \frac{e^{-x^{2}/2}}{\sqrt{2\pi}}\left(u''(x)^2-u'(x)^2\right).
\end{equation}
We see that the complexity is second order in $\kappa$ and has two distinct solution branches. To leading order, the complexity is zero in the first branch, and is $\Theta\left[u\right]$ in the other branch. The total complexity of the system is the maximum of these two. The potential $u(x)$ dictates the sign of $\Theta[u]$ and consequently which branch dominates. We show that $\Theta\left[u\right]>0$ for any potential with $u'(0)=0$ i.e. without an external field component (see SI). Thus, addition of a weak generic anharmonic potential brings about an exponential number of critical points for any non-zero magnitude (a phase transition at zero). Also, we see that the relation of the complexity with the nonlinearity strength does not depend on the details of the potential but only on the number $\Theta\left[u\right]$. We defer the discussion on the case $\Theta\left[u\right]<0$ to the part on response to an external field.

\begin{figure}
\includegraphics[scale=0.50]{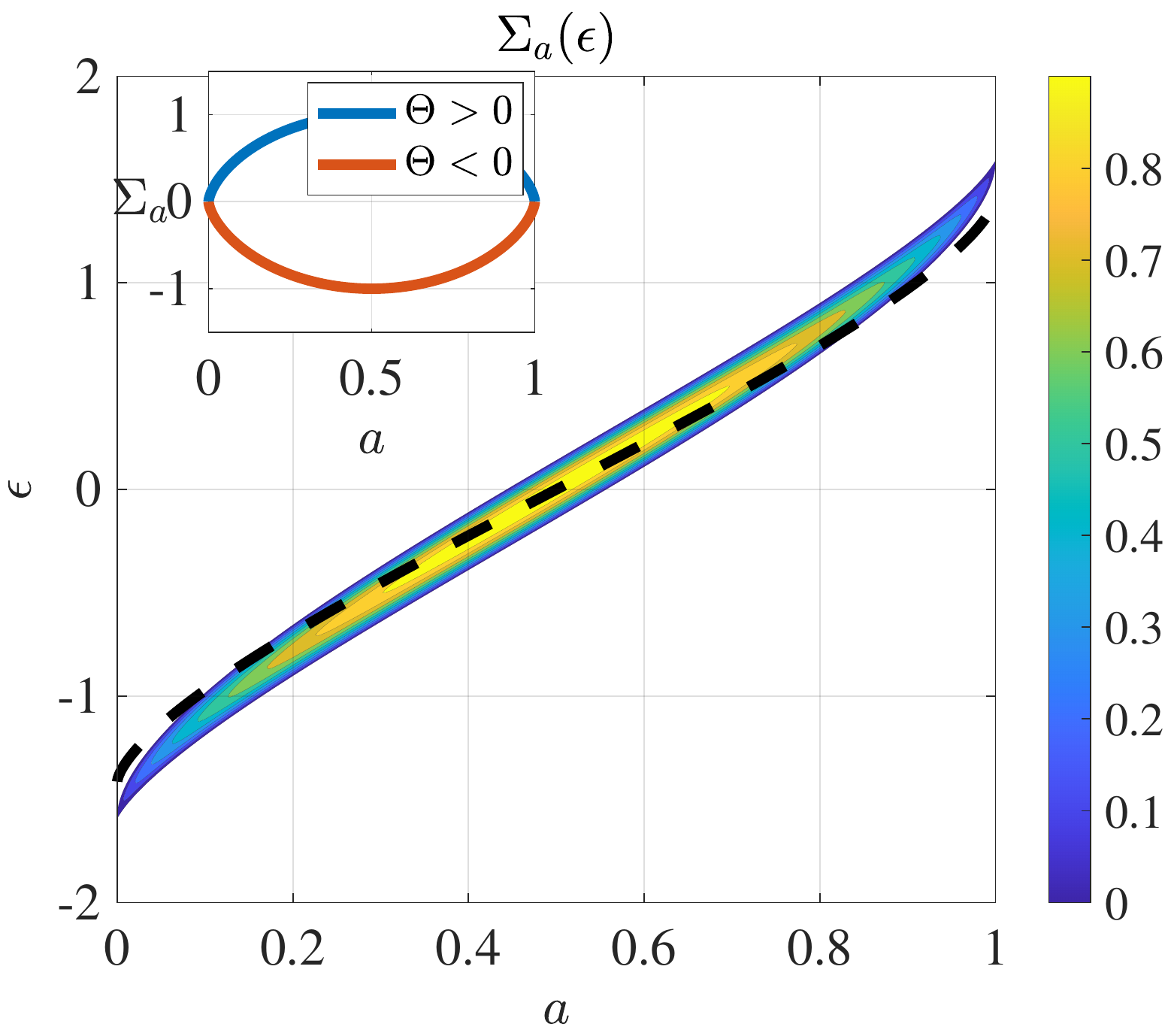} 
\caption{\label{fig:complexity_vs_index} In color: the complexity of critical points of a given index and energy $\Sigma_{a}(\epsilon)$ as a function of the normalized index $a = k/N$ and energy $\epsilon = E/N$ for $\Theta=1$, $\tilde{\Delta}=3$ (corresponding to $u(x)=\frac{4}{9}x^3$) and $\kappa = 0.2$. Dashed line: Relation between $\epsilon$ and $a$ in the unperturbed $2$-spin model. Inset: The complexity $\Sigma_{a}$ for a given index $k=a N$. The values are normalized to the maximal value. In blue, $\Sigma_{a}$ is the plotted vs. $a$ in the case where $\Theta[u]$ is positive. In red, we depict the case where $\Theta[u]$ is negative.}
\end{figure}

Next, to gain further insight about the geometry of the landscape of our model we calculate the complexity of critical points with a given index $k=a N$ and a given energy $E = \epsilon N$. This is found to be (see SI):
\begin{equation} \label{eq:complexity_index_energy}
\begin{split}
    & \Sigma_{a}(\epsilon)[u] = \kappa^{2}\left(1 - \frac{\eta^{2}(a)}{4}\right)\Theta[u]
    \\
    & -\frac{\kappa^{2}}{2}\left(\frac{1}{\kappa^{2}}\left(\eta(a)  -\epsilon\sqrt{2}\right) + \eta(a)\tilde{\Delta} \right)^{2} + O\left(\kappa^{3}\right)
\end{split}
\end{equation}
where $\tilde{\Delta}$ is a functional of $u(x)$ which is defined in the SI. $\eta(a)$ is the solution of:
\begin{equation}
    \frac{1}{2\pi}\int_{-2}^{\eta(a)} dx \sqrt{4-x^{2}}  = a.
\end{equation}
Note that this expression is valid when $\eta(a) - \epsilon\sqrt{2}=O(\kappa^{2})$ where outside this range the complexity is highly negative i.e. the average number of critical points in exponentially negligible (see SI). The dependence of $\Sigma_{a}(\epsilon)$, to leading order in $\kappa$, on the index $a$ and the energy $\epsilon$ is plotted in Fig.(\ref{fig:complexity_vs_index}) where it is seen that critical points with positive complexity are narrowly concentrated around a line in the $a \epsilon$ plane defined by 
\begin{equation} \label{eq:index_energy_concentraion}
    \epsilon = \frac{\eta(a)}{\sqrt{2}}(1 + \kappa^{2}\tilde{\Delta}).
\end{equation}
This tight connection between energy and index is also found in other models \cite{Bray2007StatisticsSpaces,Baity-Jesi2019ComparingSystems,Auffinger2013ComplexitySphere} as well as in the unperturbed model. However, in the unperturbed case only critical points obeying $\epsilon = \eta(a)/\sqrt{2}$ exist and the probability to find other critical points vanishes like $O(\exp(-N^2))$ \cite{Auffinger2013RandomGlasses}. Thus the on-site potential serves to broaden the distribution of critical points to include an exponential number of critical points outside the above relation. We also see that the on-site potential shifts by $O(\kappa^{2})$ the concentration line of critical points. This broadening and shift is similar to what occurs in the transition from pure $p$-spin models to mixed $p$-spin models. This is expected to have a profound effect on the relaxation dynamics as described in \cite{Folena2020}.
In addition, we see that critical points of index $k=N/2$ and zero energy have the highest complexity. The marginal distribution of the complexity vs the index is depicted in the inset of Fig. \ref{fig:complexity_vs_index}. We see that saddle points of extensive index become exponentially abundant while the number of critical points with $\lim_{N\rightarrow\infty}k/N = 0$ does not change on the exponential scale. Moreover, we show in the SI that the number of maxima and minima does not change as well \ref{Complexity_index_p_2}. This increase in saddles makes the landscape between minima rugged with additional barriers \cite{Ros2019ComplexitySystems}. This is expected to have implications on the lifetime of metastable states as well as relaxation pathways \cite{Rizzo2021}. We also note that the distribution of critical points in the regime of weak non-linearity is universal in that it does not depend on the details of $u(x)$ but only on ``macroscopic'' functionals of $u(x)$ ($\Theta$ and $\Tilde{\Delta}$).

As an illustrating example, let us consider the specific example of a network of coupled optical oscillators (lasers \cite{Nixon2013}  or DOPOs \cite{Wang2013}). In these systems the coupling between oscillators is typically bi-linear while the on-site potential's form and strength depend on an external driving (pumping): For zero or weak driving the on-site potential is harmonic (and thus trivial) while increasing the drive adds an an-harmonic term to the potential which can lead to a pitchfork bifurcation resulting in bi-stability. The experimental realization of large systems of this sort and their use to simulate spin models as well as heuristic machines for optimization has led to an interest in their dynamics in various regimes \cite{Tatsumura2021,inagaki2016large,bohm2018understanding}. Our results indicate that the energy landscape of these systems becomes rugged and complex even slightly above the system's oscillation threshold where the non-linearity is weak. This means that the dynamics of these systems are expected to be nontrivial in this regime as well and to become more so as the non-linearity is further increased.


\begin{figure}
\includegraphics[scale=0.5]{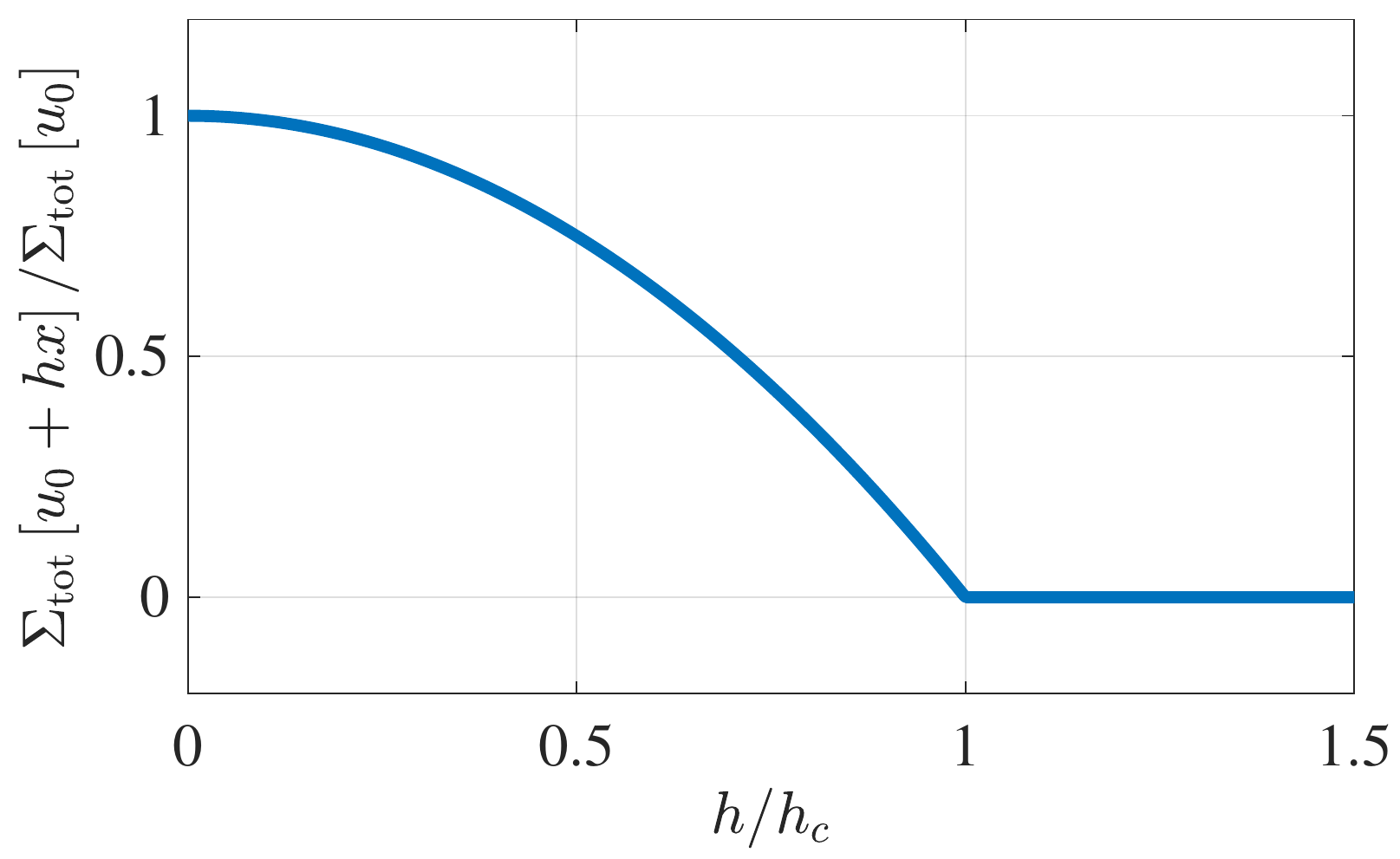}
\caption{\label{fig:transition_example} The complexity $\Sigma_{\mathrm{tot}}$ second order correction as a function of the magnetic field strength, scaled by the critical strength, for the case of an on-site potential with $\overline{u_{0}'} = 0$}
\end{figure}

\paragraph*{Response to External Field} As shown in appendix D of the SI, a strong enough external field $h = u'(0)$ leads to negative $\Theta[u]$ (see \eqref{eq:Theta_def}). In this case the complexity is zero to leading order and our results are similar to those in \cite{Belius2022TrivialityField, Fyodorov2014TopologyOptimization} where ``trivial topology'' of the landscape was found under a sufficiently strong external magnetic field. In the trivial topology phase, the energy landscape has only two critical points (a minimum and a maximum) \cite{Belius2022TrivialityField}. Our results for the complexity of a given index $\Sigma_{a}$ as depicted in the inset of Fig. \ref{fig:complexity_vs_index}, in this case, are consistent with this picture: the complexity is non-negative only for indices $k=o(N)$ (or $k=N-o(N)$), which represent critical points that are a minimum (maximum) in most directions. From the perspective of the spin value distribution, we show (see eq. (59) in SI) that the distribution is skewed in the direction of the applied magnetic field, s.t. for $u(x) = h x$ this leads to an $O(\kappa)$ shift in the disorder averaged magnetization $\overline{\sigma}  \sim \kappa h$. This shift directly shows the alignment of the spins along the external field at the maxima or minima of the model. 

Having explored both system states, we now turn to discussing the transition between them. As shown in \eqref{eq:power_series_Theta}, driving the system between the states can be achieved by changing the magnetic field strength $u'(0)$. Assume that for a specific potential $u_{0}(x)$ the added complexity $\Theta \left[u_{0}\right]$ is positive, e.g. any potential with $u_{0}'(0) = 0$. Next, we add an external field of strength $h$ s.t. $u(x) = u_{0}(x) + h x$. A simple calculation shows:
\begin{equation}
\begin{split}
     \Theta\left[u\right] = 
     \Theta\left[u_{0}\right] - \frac{h^{2}}{4} - \frac{h}{2}\overline{u'}.
\end{split}
\end{equation}
where $\overline{u(x)}$ denotes averaging the weighted average of $u(x)$ w.r.t the standard Gaussian distribution. It is seen that for large enough field strength, i.e. $h>h_{c+}$ or $h<h_{c-}$, $\Theta \left[u\right]$ drops below zero. At $h_c$ there is a transition to the zero complexity branch.  $h_c$ is  found by equating the above expression to zero:
\begin{equation}
\begin{split}
    h_{\mathrm{c\pm}} = \pm \sqrt{\overline{u'}+ 4\Theta\left[u_{0}\right]}-\overline{u'}
\end{split}
\end{equation}
It is also seen that the form of this transition is independent of the form of $u_{0}(x)$.

Although our main motivation is to analyze perturbations of the 2-spin spherical model, the same analysis can be readily applied for a general $p$-spin spherical model. The average total complexity $\Sigma^{(p>2)}_{\mathrm{tot}}$ reads in that case
\begin{equation}
    \Sigma^{(p>2)}_{\mathrm{tot}}[u] = \frac{1}{2}\log \left(p-1\right) +  \kappa^{2}\Theta^{(p)}\left[u\right] + O(\kappa^{3})
\end{equation}
where $\Theta^{(p)}\left[u\right]$ appears in Eq. (9) of the SI.
Note that the unperturbed complexity is consistent with \cite{Auffinger2013RandomGlasses}. As in the 2-spin spherical model, the correction to the total complexity for the $p$-spin model is $O(\kappa^{2})$. However, there is only a single solution branch and  the total complexity is always positive (in the regime of weak $\kappa$). Moreover, the on-site potential can either increase or decrease the total complexity depending on its form, in contrast to the 2-spin spherical model, where it can only be increased. Specifically, we also recover the results of \cite{Belius2022TrivialityField} in the regime of small external magnetic field. In Eq. (9) of the SI we also provide an expression for the complexity $\Sigma^{(p>2)}_{a}[u]$ of the $p$-spin model with extensive fixed index $k=N a$ for any integer $p$.

\paragraph*{Derivation Outline}
The full derivation of our results is quite lengthy and technical. We therefore provide only the outline of the calculation of the complexity defined in \eqref{eq:Def_Complexity} for the $p=2$ case. The full derivation for this and other cases is given in the SI. We start the calculation of \eqref{eq:Kac_Rice} by evaluating the disorder average $\overline{ \left| \det\left( \nabla^{2}H \right) \right| \delta\left( \nabla H \right)}$ at a fixed position $\sigma$ on the sphere. The average is taken over the joint distribution of the covariant Hessian  and gradient of $\phi(\sigma)$. First, we use the independence of $\nabla\phi$ and $\nabla\phi^{2}$ to factor out the expectation as follows 
\begin{equation} \label{eq:disorder_E_factor}
\begin{split}
    & \overline{\left| \det\left( \nabla^{2}H \right) \right| \delta\left( \nabla H \right)}  = \overline{\delta\left( \nabla H \right)} \, \, \overline{\left| \det\left( \nabla^{2}H \right) \right|}.
\end{split}
\end{equation}
To evaluate the expectation values in Eq. \eqref{eq:disorder_E_factor}  we start by writing the distribution of the random matrix $\nabla^{2}\phi$ and $\nabla\phi$ (see SI appendix A). Equipped with this pdf, the integral $\overline{ \delta\left( \nabla H \right)}$ is easily carried out to give:
\begin{equation}
    \overline{\delta\left( \nabla H(\sigma) \right)} = \left(\frac{1}{2\pi p}\right)^{N/2}\exp\left(-\frac{\kappa^{2}}{2 p} \lVert\nabla U\left(\sigma\right)\rVert^{2}\right).
\end{equation}
The evaluation of $\overline{\left|\mathrm{det} \nabla^{2}H\left(\sigma\right)\right|}$ is more involved and constitutes the main technical contribution of this work. The challenge is due to $\nabla^{2}H$ being a sum of non-commuting matrices: the random matrix $\nabla^{2}\phi$ and the diagonal deterministic matrix $\kappa\nabla^{2}U$ (see sup. mat. for explicit expressions). We address this difficulty by resorting to a perturbative evaluation of the eigenvalue distribution of $\nabla^{2}H$ in the small $\kappa$ regime. This distribution is used in turn to calculate the determinant modulus as described below.  

To evaluate $\overline{\left|\mathrm{det} \nabla^{2}H\left(\sigma\right)\right|}$ we use recent results (Proposition 5.3 in \cite{Arous2021LandscapeManifold}, Theorem 4.1 in \cite{BenArous2021ExponentialInvariance}),
to formally express the expectation value in terms of the eigenvalue distributions of $\nabla^{2}\phi$ and $\kappa\nabla^{2}U$,
\begin{equation} \label{eq:hessian_determinant_expectation_2}
\begin{split}
    & \lim_{N\to\infty}\frac{1}{N}\log\overline{\left|\mathrm{det} \nabla^{2}H\left(\sigma\right)\right|}  =
    \\
    & \sup_{s} \int d\lambda \, \nu_{\nabla^{2}H(\sigma)}\left(\lambda\right) \log\left|\lambda-s\right| - \frac{s^{2}}{2 p^{2}},
\end{split}
\end{equation}
where $\nu_{\nabla^{2}H(\sigma)}$ is the eigenvalue distribution of $\nabla^{2}H\left(\sigma\right)$. This distribution, in the limit of large $N$, converges to the free convolution, denoted $\boxplus$, of the eigenvalue distributions of $\nabla^{2} \phi$ and $\kappa\nabla^{2}U$ \cite{ Anderson2009AnMatrices, Arous2021LandscapeManifold},
\begin{equation}
    \nu_{\nabla^{2}H}\left(\lambda\right) = \left(\rho_{\mathrm{sc}}\boxplus \nu_{\kappa\nabla^{2}U}\right) \left(\lambda\right),
\end{equation}
where $\nu_{\kappa\nabla^{2}U}(\lambda)$ is the eigenvalue distribution of $\kappa\nabla^{2}U$ and $\rho_{\mathrm{sc}}(\lambda)$ is the Wigner semicircle distribution with variance $V_{p}\equiv p(p-1)$:
\begin{equation}
    \rho_{\mathrm{sc}}\left(\lambda\right) = \frac{1}{2\pi V_{p}}\sqrt{4V_{p}-\lambda^{2}}.
\end{equation}
Intuitively, the free convolution is the random matrix analogue to the convolution operation used to calculate the probability density of a sum of independent random variables. It is non-linear and generally the result cannot be expressed in a closed form. However, in the regime of small $\kappa$ we are able to find a perturbative expression for $\nu_{\nabla^{2}H}\left(\lambda\right)$ by solving perturbatively the Pastur integral equation \cite{Biane1997OnDistribution}. This results, up to $O(\kappa^{3}),$  is a shifted and widened semi-circular distribution given by:
\begin{align}
    &\nu_{\nabla^{2}H}\left(\lambda\right) =\\ &\frac{1}{\sqrt{1+\kappa^{2}V_{\nabla^{2}U}}}\rho_{\mathrm{sc}}\left(\frac{\lambda-\kappa \sqrt{V_{p}}m_{\nabla^{2}U}}{\sqrt{1+\kappa^{2}V_{\nabla^{2}U}}}\right)+O(\kappa^3)\;,\nonumber
\end{align}
with $m_{\nabla^{2}U}$ and $V_{\nabla^{2}U}$ the mean and variance of the eigenvalue distribution of $\frac{1}{\sqrt{V_{p}}}\nabla^{2}U$, respectively. Now we can proceed to derive a closed form expression for $\overline{ \left| \det\left( \nabla^{2}H \right) \right|}$ by substitution of the above expression for $\nu_{\nabla^{2}H}\left(\lambda\right)$ in Eq. \ref{eq:hessian_determinant_expectation_2} and optimization w.r.t $s$:
\begin{equation}
\begin{split}
    & \lim_{N\rightarrow\infty} \frac{1}{N}\log\overline{\left|\mathrm{det} \nabla^{2}H\left(\sigma\right)\right|} = \frac{1}{2}\left(-1+\log 2\right) -
    \\
    &  \frac{1}{4}\kappa^{2}m_{\nabla^{2}U}^{2} + \kappa^{2}V_{\nabla^{2}U} 
    \begin{cases} 
        \left|\eta_{2}\right| - \frac{1}{2} \, &  \left|\eta_{2}(\sigma)\right| \ge 2 \\
        \frac{1}{2} + \frac{1}{4}\eta_{2}^{2} &  \left|\eta_{2}(\sigma)\right|<2
    \end{cases},
\end{split}
\end{equation}
where we ignored terms which are $O(\kappa)^3$ and $\eta_{2} = \frac{m_{\nabla^{2}U}/\kappa}{V_{\nabla^{2}U}}$. Note that the result contains two branches. The form of the supremum equation and its solution branches is similar to the one found in \cite{Belius2022TrivialityField} as was discussed in detail following the results statement in the previous section.

Now we turn to the evaluation of the integration over the $N$-sphere in \eqref{eq:Kac_Rice}: 
\begin{equation}
    \overline{\mathcal{N}_{\mathrm{tot}}}   = \left(\frac{1}{2\pi p}\right)^{\frac{N}{2}} \int_{\mathbb{S}^{N-1}} d\sigma \exp\left(-\frac{\kappa^{2}}{2 p} \lVert\nabla U\rVert^{2}\right)  \overline{\left|\mathrm{det}\nabla^{2}H\right|} .
\end{equation}
The approach we adopt for the evaluation of this integral is to pass from spatial integration over the surface of the sphere to functional integration over the empirical distribution of $\sigma_{i}$ (the "Coulomb gas" technique  \cite{Dyson1962StatisticalI}). This results in a functional optimization problem whose formulation and perturbative solution are described in detail in the SI sec. \ref{sec:sphere_functional_integration}).

\paragraph*{Discussion}
We found that introducing an ``on-site'' non-linearity to a disordered mean-field coupled system can significantly increase the complexity of the energy landscape of the system. We characterised the changes to the geometry of the energy landscape through the distribution of critical points w.r.t energy and index and found them to exhibit a universal behaviour independent of the specifics of the added non-linearity. These results are immediately applicable to the characterisation of the energy landscape of physical systems with ``on-site'' weak non-linearity and disordered long range coupling such as ``soft-spin'' glass models \cite{Castellani2005Spin-glassPedestrians, Cavagna1998StationaryEnergy, Cavagna2001RoleTransition, mezard1987spin}, coupled oscillator networks \cite{Nixon2013, Wang2013,Inagaki2016} and atomic spin gases \cite{Horowicz2021CriticalGas}.

Future research directions include extending the results of this work beyond the perturbative regime, relaxation of the spherical constraint to study more diverse physical models, studying the quenched complexity, possibly by carrying out a second moment analysis \cite{Subag2017TheApproach}. Also, it is of interest to consider short range coupling schemes possibly by considering banded random matrix models \cite{Arous2021LandscapeManifold, Bourgade2017UniversalityMatrices}.

\paragraph*{Acknowledgments}
O. R. is the incumbent of the Shlomo and Michla Tomarin career development chair, and is supported by the Abramson Family Center for Young Scientists, the Israel Science Foundation Grant No. 950/19 and by the Minerva foundation. IG is grateful to Amy Entin for her help with writing this manuscript. B. LACT is supported by the EPSRC Grant EP/V002473/1 Random Hessians and Jacobians:
theory and applications. O. Z. was partially supported by the European Research Council (ERC) under the European Union's Horizon 2020 research and innovation programme (grant agreement No. 692452) and by the Israel Science Foundation grant 421/20. E. S. is the incumbent of the Skirball Chair in New Scientists, and is supported by the Israel Science Foundation grant 2055/21.

\bibliography{references}

\widetext

\newpage


\begin{center}
\textbf{\Large Supplemental Information: On-Site Potential Creates Complexity in Systems with Disordered Coupling}
\end{center}

\setcounter{equation}{0}
\setcounter{figure}{0}
\setcounter{table}{0}
\setcounter{page}{1}
\makeatletter
\renewcommand{\theequation}{S\arabic{equation}}
\renewcommand{\thefigure}{S\arabic{figure}}

\subfile{SI}


\end{document}

%% file: SI.tex
\preprint{}

\title{Supplementary Information: On-Site Potential Creates Complexity in Systems with Disordered Coupling}

\maketitle
\widetext


\part{Setup}
Define an ``energy landscape" on the $N$-sphere composed of a ``pure" $p$-spin Gaussian field $\phi(\sigma)$  and an ``on-site" deterministic potential $U(\sigma)$:
\begin{equation}
    H\left(\sigma\right) = \phi\left(\sigma\right) + \kappa U\left(\sigma\right) = \phi\left(\sigma\right) + \kappa\sum_{i=1}^{N} u\left(\sigma_{i}\right)
\end{equation}
where $\sigma \in\sqrt{N}\mathbb{S}^{N-1}$,  $\kappa$ is the ``on-site" potential strength parameter, and $u(x)$ is the potential function, assumed throughout this work to be smooth and of bounded derivatives.  The Gaussian field on the sphere is defined by the following correlation function corresponding to the ``pure" p-spin spherical model:  
\begin{equation}
    \overline{\phi\left(\sigma\right)\phi\left(\sigma'\right)} = N \left(\frac{\sigma \cdot \sigma'}{N}\right)^{p} \, .
\end{equation}
To evaluate the average total number of critical points on the sphere, we use the spherical analogue of the Kac-Rice formula (as given in \cite{Auffinger2013RandomGlasses})
\begin{equation} \label{eq:fixed_point_num_def}
    \overline{ \mathcal{N}_{\mathrm{tot}} } = \int_{\sqrt{N}\mathbb{S}^{N-1}} d\sigma_{N-1} \overline{ \left| \det\left( \nabla^{2}H \right) \right| \delta\left( \nabla H \right)}
    \end{equation}
where $\nabla H$ and $\nabla^{2}H$ are respectively the covariant derivative and Hessian matrix of $H$ on the sphere, and $\overline{X}$ denotes averaging a quantity $X$ over the random field distribution.
Defining the index as the number of negative eigenvalues of the Hessian, the average number of critical points of index $k$ is similarly given by
\begin{equation} \label{eq:fixed_point_k_num_def}
    \overline{ \mathcal{N}_{k} } = \int_{\sqrt{N}\mathbb{S}^{N-1}} d\sigma_{N-1} \overline{ \left| \det\left( \nabla^{2}H \right) \right| \delta\left( \nabla H \right) \delta\left( I\left(\nabla^{2} H\right) - k \right)}\;,
\end{equation}
where $I(\cdot)$ is the index function; see \cite{Bray2007StatisticsSpaces}. In a similar manner, we define the average number of critical points of index $k$ and energy $E$
\begin{equation} \label{eq:fixed_point_k_E_num_def}
    \overline{ \mathcal{N}_{k} (E) } = \int_{\sqrt{N}\mathbb{S}^{N-1}} d\sigma_{N-1} \overline{ \left| \det\left( \nabla^{2}H \right) \right| \delta\left( \nabla H \right) \delta\left( I\left(\nabla^{2} H\right) - k \right)\delta\left(H-E\right)}\;,
\end{equation}
In this work, we are interested in the annealed (that is, averaged over the Gaussian field $\phi$) complexity of the number of critical points. The total complexity and the complexity of critical points of extensive index $k = \lfloor a N\rfloor$ 
(with $a \in \left(0,1\right)$) are defined respectively as:
\begin{align}
    \Sigma_{\mathrm{tot}} = \Sigma_{\mathrm{tot}}\left[u(x) \right] = & \lim_{N\rightarrow\infty} \frac{1}{N} \log \overline{ \mathcal{N}_{\mathrm{tot}} } 
    \\
    \Sigma_{a} = \Sigma_{a}\left[u(x) \right] = & \lim_{N\rightarrow\infty} \frac{1}{N} \log \overline{ \mathcal{N}_{a N} }
    \\
    \Sigma_{a}(E) = \Sigma_{a}\left[u(x) \right] = & \lim_{N\rightarrow\infty} \frac{1}{N} \log \overline{ \mathcal{N}_{a N}(E) }
\end{align}
where the limit in the second and third equations is taken such that $a$ is fixed. 

\part{Main Results}
Our main results are given by the statements in Sections \ref{sec-I}, \ref{sec-II}  and \ref{sec-IIa}. The derivation of these results are in the following parts.
\section{Complexity of the Mean Total Number of Critical Points}
\label{sec-I}
The total annealed complexity of a $p$-spin model perturbed by a weak ``on-site" potential, $\Sigma_{\mathrm{tot}}^{(p)}$, as a functional of the potential $u(\cdot)$, is given by:
\begin{align}
    \Sigma^{(p=2)}_{\mathrm{tot}}[u] & =  \kappa^{2}\max\left(0,\Theta^{(2)}\left[u\right]\right) + O(\kappa^{3})
    \\
    \Sigma^{(p>2)}_{\mathrm{tot}}[u] & = \frac{1}{2}\log \left(p-1\right) +  \kappa^{2}\Theta^{(p)}\left[u\right] + O(\kappa^{3})
\end{align}
where
\begin{align}
\label{eq-Thetag}
    \Theta^{(p)}\left[u\right] =& \frac{1}{2p}\int dx \frac{e^{-x^{2}/2}}{\sqrt{2\pi}}\left(\frac{1}{p-1}u''(x)^2-u'(x)^2\right)\\
    &+ \frac{p-2}{2p(p-1)}\left(\int \frac{e^{-x^{2}/2}}{\sqrt{2\pi}} u''(x)  dx\right)^{2}.\nonumber
\end{align}
Note that for $p=2$, Eq. \ref{eq-Thetag} specializes to
\begin{equation}
    \Theta^{(p=2)}\left[u\right] = \frac{1}{4}\int dx \frac{e^{-x^{2}/2}}{\sqrt{2\pi}}\left(u''(x)^2-u'(x)^2\right).
\end{equation}
\section{Complexity of the Mean Number of Critical Points of a Given Index}
\label{sec-II}
\begin{figure}
\includegraphics[scale=0.50]{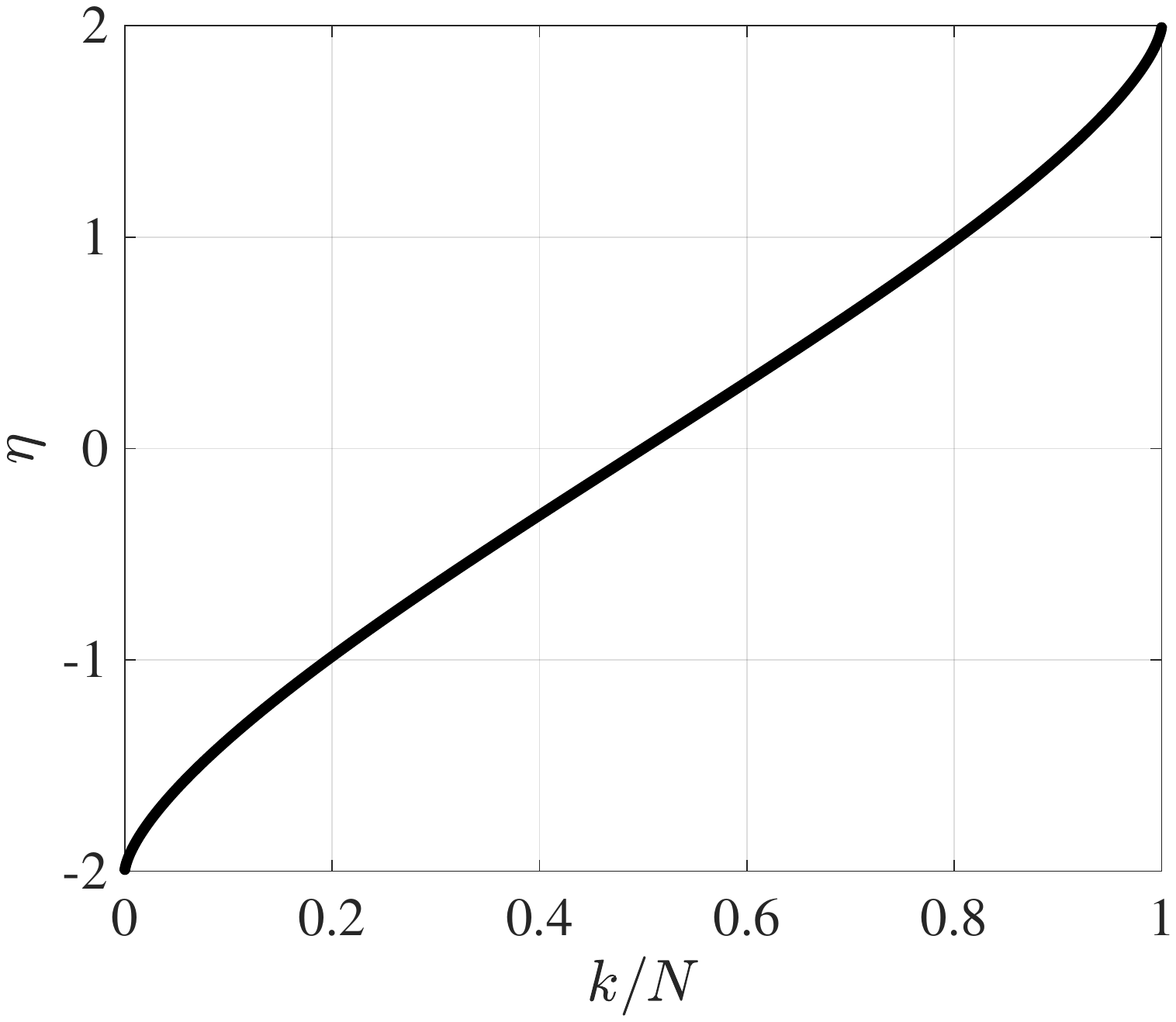} 
\caption{\label{fig:eta_vs_index} Plot of $\eta(a)$}
\end{figure}

The annealed complexity of critical points of (extensive) index $k = \lfloor a N\rfloor$ of a $p$-spin model perturbed by a weak ``on-site" potential, $\Sigma_{a}^{(p)}$, as a functional of the potential $u$, is given by
\begin{equation}
    \Sigma^{(p)}_{a}[u] = \frac{1}{2}\log \left(p-1\right) -\frac{p-2}{4p}\eta^{2}(a) +
    \kappa^{2}\left(1 - \eta^{2}(a)\frac{p-1}{p^{2}}\right)\Theta^{(p)}[u] + O\left(\kappa^{3}\right)
\end{equation}
where $\eta(a)\in\left(-2,2\right)$ is defined by (see figure \ref{fig:eta_vs_index})
\begin{equation}
    \frac{1}{2\pi}\int_{-2}^{\eta(a)}  \sqrt{4-x^{2}} \, dx = a.
\end{equation}

\section{Complexity of the Mean Number of Critical Points of a Given Index and Energy}
\label{sec-IIa}
The annealed complexity of critical points of (extensive) index $k = \lfloor a N\rfloor$ and (extensive) energy $E=\epsilon N$ of a $p$-spin model perturbed by a weak ``on-site" potential, $\Sigma_{a}^{(p)}(\epsilon)$, as a functional of the potential $u$, in the regime where $\eta(a) - \frac{\epsilon}{\sqrt{p(p-1)}} = O(\kappa^{2})$ is given by
\begin{equation}
    \Sigma_a^{(p)}(\epsilon) = \Sigma_a^{(p)} - \frac{\kappa^{2}}{2}\left(\frac{1}{\kappa^{2}}\left(\eta(a) - \epsilon\sqrt{\frac{p}{p-1}}\right) + \eta(a)\tilde{\Delta} \right)^{2} + O(\kappa^{3})
\end{equation}
where $\Sigma_a^{(p)}$ is the complexity for a given index, $\eta(a)\in\left(-2,2\right)$ and $\tilde{\Delta}$ are defined below
\begin{equation}
    \frac{1}{2\pi}\int_{-2}^{\eta(a)}  \sqrt{4-x^{2}} \, dx = a.
\end{equation}
\begin{equation}
    \tilde{\Delta} = \left(\frac{1}{2 p(p-1)} - \frac{1}{p^{2}}\right) \overline{u''^{2}} + \left(-\frac{1}{p^{2}} + \frac{1}{p}\right)\overline{u'^{2}}
\end{equation}
where here the $\overline{f(x)}$ notation denotes averaging w.r.t the standard Gaussian measure. The calculation of this complexity outside the regime $\eta(a) - \sqrt{\frac{p}{p-1}}\epsilon = O(\kappa^{2})$ is beyond the scope of this work.

\part{Results Derivation}
\section{Evaluation of the Mean Total Number of Critical Points}
\subsection{Preliminaries}
We start the calculation of \eqref{eq:fixed_point_num_def} by evaluating the disorder average $\overline{ \left| \det\left( \nabla^{2}H(\sigma) \right) \right| \delta\left( \nabla H(\sigma) \right) }$ at a fixed position $\sigma$ on the sphere. The average is taken over the joint distribution of the covariant Hessian  and gradient of the field $\phi$. See Appendix \ref{Appendix A: Dist of Random Fields} for detailed description of the joint distribution of the fields $\phi,\, \nabla\phi,\, \nabla^{2}\phi$ adopted from \cite{Auffinger2013RandomGlasses}. We note here, for the sake of clarity, the main properties of this distribution which are instrumental for our calculation:
\begin{enumerate}
    \item At each point $\sigma$, the field $\phi$ and its derivatives 
    are mean zero jointly normally distributed random variables.
    \item 
    At any point $\sigma$,  $\phi(\sigma)$ and $\nabla^{2}\phi(\sigma)$ are independent of $\nabla \phi(\sigma)$. \label{item:gradient_component_indepedence}
    \item The components of $\nabla\phi(\sigma)$ are independent. The variance of each component is $\mathrm{Var}\left(\left(\nabla\phi\right)_m\right) = p$.
    \item The random matrix $\nabla^{2}\phi$ has the same distribution as $\sqrt{p(p-1)}M + \xi p\mathrm{I}_{N-1}$ where $M\sim \sqrt{\frac{N-1}{N}}\mathrm{GOE}_{N-1}$, $\xi\sim \mathrm{N}\left(0,\frac{1}{N}\right)$, $\mathrm{I}_{N-1}$ is the identity matrix of size $(N-1)\times(N-1)$. We denote by $\mathrm{GOE}_{n}$ the Gaussian orthogonal ensemble of dimension $n$ \cite{Anderson2009AnMatrices} with an off-diagonal variance of $\frac{1}{n}$ i.e. if $A \sim GOE_{n}$ then $\mathbb{E}\left[A_{ij}^{2}\right] = \frac{1+\delta_{ij}}{n}$. \label{item:Random_Hessian_distribution}
    \item The conditional distribution of the random matrix $\nabla^{2}\phi$, given a value $E$ for the random field, $\phi$ is $\sqrt{p(p-1)}M -E p\mathrm{I}_{N-1}$ where $M\sim \sqrt{\frac{N-1}{N}}\mathrm{GOE}_{N-1}$. \label{item:Random_Hessian_conditional_distribution}
\end{enumerate}

First, we use the independence of $\nabla\phi$ and $\nabla\phi^{2}$ to factorize the expectation in \eqref{eq:fixed_point_num_def}
as follows: 
\begin{equation}
\begin{split}
    & \overline{ \left| \det\left( \nabla^{2}H \right) \right| \delta\left( \nabla H \right) }  = \overline{ \delta\left( \nabla H \right)  } \times \overline{ \left| \det\left( \nabla^{2}H \right) \right|}\,.
\end{split}
\end{equation}
The average $\overline{ \delta\left( \nabla H(\sigma) \right) }$ is easily carried out and gives:
\begin{equation}
\begin{split}
    \overline{ \delta\left( \nabla H(\sigma) \right) } = & \prod_{m=1}^{N-1} \overline{ \delta\left( \left(\nabla \phi(\sigma)\right)_{m} + \kappa\left(\nabla U(\sigma)\right)_{m}\right) } = \prod_{m=1}^{N-1} \sqrt{\frac{1}{2\pi p}}\exp\left(-\frac{\kappa^{2}}{2 p} \left(\nabla U\left(\sigma\right)\right)_{m}^{2}\right)
    \\
    =& \left(\frac{1}{2\pi p}\right)^{(N-1)/2}\exp\left(-\frac{\kappa^{2}}{2 p} \lVert\nabla U\left(\sigma\right)\rVert^{2}\right) ,
\end{split}
\end{equation}
where in the first equality we used the gradient component independence and in the second the normal distribution of each component.

We evaluate $\overline{ | \det\left( \nabla^{2}H \right)| }$ by using the distribution in \ref{item:Random_Hessian_distribution} as follows \cite{Auffinger2013RandomGlasses}:
\begin{equation}
\begin{split}
     \overline{ \left| \det\left( \nabla^{2}H \right) \right|} = & \overline{\left |  \det\left( \nabla^{2} \phi + \kappa\nabla^{2}U\right)\right| } 
     \\
     = & \sqrt{\frac{N}{2\pi p^{2}}}\int_{\mathbb{R}}\exp{\left(-N\frac{s^{2}}{2p^{2}}\right)}\overline{ \left|\mathrm{det} \sqrt{p(p-1)}M + \kappa\nabla^{2}U + s\mathrm{I}_{N-1}\right| }\, ds \; ,
\end{split}
\end{equation}
where $M\sim\sqrt{\frac{N-1}{N}}\mathrm{GOE}_{N-1}$ and it is understood that the averaging is carried out over the matrix $M$. In the second equality we have used the distribution of $\nabla^{2}\phi$ defined in item \ref{item:Random_Hessian_distribution} above. Next, we employ the same method as in Proposition 5.3 of \cite{Arous2021LandscapeManifold} and use Theorem 4.1 in \cite{BenArous2021ExponentialInvariance} to get
\begin{equation} \label{eq:sup_mean_abs_det}
     \lim_{N\to \infty}\frac{1}{N}\log \left(\overline{\left|\mathrm{det} \nabla^{2}H\left(\sigma\right)\right|} \right)  = \sup_{s} \left(\int d\lambda \, \left(\rho_{\mathrm{sc}}\boxplus \nu_{\kappa\nabla^{2}U}\right) \left(\lambda\right) \log\left|\lambda+s\right| - \frac{s^{2}}{2 p^{2}}\right) \, ,
\end{equation}
where $\boxplus$ is the free-convolution \cite{Anderson2009AnMatrices},   $\rho_{\mathrm{sc}}$ is the Wigner semicircle of variance $p(p-1)$ and $\nu_{\kappa\nabla^{2}U}$ is the large $N$ limit of the empirical spectral distribution of $\kappa\nabla^{2}U$, i.e.
\begin{align}
    \rho_{\mathrm{sc}}\left(\lambda\right) = & \frac{1}{2\pi p\left(p-1\right)}\sqrt{4p\left(p-1\right)-\lambda^{2}} \, \Theta_{H}\left(2\sqrt{p(p-1)}-|\lambda|\right)
\\
    \nu_{\kappa\nabla^{2}U}\left(\lambda\right) = & \lim_{N\rightarrow\infty}\frac{1}{N-1}\sum_{m=1}^{N-1}\delta\left(\lambda - \kappa\lambda_m \left(\nabla^{2}U \right) \right)
\end{align}
where $\Theta_{H}$ denotes the Heaviside step function and $\lambda_m\left(\nabla^{2}U \right)$ with  $m=1,\ldots,N-1$ are the eigenvalues of the matrix $\nabla^{2}U$.
\subsection{Free convolution evaluation}
The free convolution operation is non-linear and the result cannot generally be expressed in a closed form. However, in the regime of small $\kappa$, we are able to solve the Pastur integral equation \cite{Biane1997OnDistribution} and find a perturbative expression for $\nu_{\nabla^{2}H}\left(\lambda\right)$. Let $G(z)$ be the Stieltjes transform \cite{Anderson2009AnMatrices} of $\rho_{\mathrm{sc}}\boxplus \nu_{\kappa\nabla^{2}U}$. Since one of the distributions in the convolution is the semicircle, one obtains a relatively simple implicit equation for $G$ (the Pastur equation) \cite{Biane1997OnDistribution}:
\begin{equation}
    G\left(z\right) = \int \frac{\nu_{\kappa\nabla^{2}U}\left(\lambda\right) d\lambda}{z - p(p-1)G\left(z\right)-\lambda} , \quad z\in \mathbf{C}_+\, .
\end{equation}
In Appendix \ref{Appendix B N-Sphere Covariant Hessian Single Particae}, the perturbative solution of this equation is detailed up to order $O\left(\kappa^{2}\right)$. Next, the inverse transform is utilized to calculate $\rho_{\mathrm{sc}}\boxplus \nu_{\kappa\nabla^{2}U}$ from $G$. Up to $O(\kappa^{3})$, this results in a shifted and widened semi-circular distribution given by
\begin{equation} \label{eq:hessian_spectral_dist}
\begin{split}
    & \nu_{\nabla^{2}H} = \left(\rho_{\mathrm{sc}}  \boxplus \nu_{\kappa\nabla^{2}U}\right) \left(\lambda\right) = \frac{\displaystyle \sqrt{-\left(\lambda-\kappa \sqrt{p(p-1)}m_{\nabla^{2}U}\right)^{2} + 4p\left(p-1\right)\left(1 + \kappa^{2}V_{\nabla^{2}U}\right)}}{\displaystyle 2\pi p\left(p-1\right)\left(1 + \kappa^{2}V_{\nabla^{2}U}\right)}+O\left(\kappa^{3}\right)\, ,
\end{split}
\end{equation}
where we have denoted the expectation  and variance of the spectrum of $\frac{1}{\sqrt{p(p-1)}}\nabla^{2}U\left(\sigma\right)$ by $m_{\nabla^{2}U}$ and $V_{\nabla^{2}U}$, respectively, i.e.,
\begin{equation}
    \begin{split}
        & m_{\nabla^{2}U}(\sigma) \equiv \frac{1}{\sqrt{p(p-1)}}\int \lambda \, \nu_{\nabla^{2}U(\sigma)}(\lambda) d\lambda 
        \\
        & V_{\nabla^{2}U}(\sigma) \equiv \frac{1}{p(p-1)}\int (\lambda-m_{\nabla^{2}U(\sigma)})^{2} \nu_{\nabla^{2}U(\sigma)}(\lambda) d\lambda \, .
    \end{split}
\end{equation}
\subsection{Evaluation of the supremum}
Equipped with an explicit, simple expression for the spectral distribution of $\nabla^{2}H(\sigma)$, we can evaluate the integral in \eqref{eq:sup_mean_abs_det} explicitly:
\begin{equation}
    \begin{split}
       \lim_{N\to \infty}\frac{1}{N}  \log\overline{\left|\mathrm{det}\nabla^{2}H\left(\sigma\right)\right|} = & \sup_{s} \Big[\Phi\left(\frac{\kappa\sqrt{p(p-1)} m_{\nabla^{2}U}+s}{\sqrt{p(p-1)(1+\kappa^2 V_{\nabla^2 U}) }}\right) + \frac{\left(\kappa \sqrt{p(p-1)}m_{\nabla^{2}U}+s\right)^{2}}{4p\left(p-1\right)\left(1 + \kappa^{2}V_{\nabla^{2}U}\right)}
        \\
        &  +\frac{1}{2}\left(-1 + \log\left(p\left(p-1\right)\left(1 + \kappa^{2}V_{\nabla^{2}U}\right)\right)\right) - \frac{s^{2}}{2 p^{2}}\Big] + O\left(\kappa^{3}\right),
    \end{split}
\end{equation}
where
\begin{align}
    \Phi(\eta) &=\int_{-2}^2\frac{d\lambda}{2\pi}\,\sqrt{4-\lambda^2}\log|\lambda-\eta|-\frac{\eta^2}{4}+\frac{1}{2}\\
    &= -\Theta_{H}\left(|\eta| - 2\right)\left(\frac{|\eta|\sqrt{\eta^{2}-4}}{4} + \log{\frac{|\eta| - \sqrt{\eta^{2}-4}}{2}}\right)\nonumber
\end{align}
and $\Theta_{H}$ denotes the Heaviside step function.

The change of variables 
\begin{equation}
    \label{eq-eta}
    \eta= - \frac{\kappa\sqrt{p(p-1)} m_{\nabla^{2}U}+s}{\sqrt{p(p-1)(1+\kappa^2 V_{\nabla^{2}U})}}
\end{equation}
and some algebra gives:
\begin{align}
\label{eq-25}
    &\lim_{N\to \infty}\frac{1}{N}\log\overline{\left|\mathrm{det}\nabla^{2}H\left(\sigma\right)\right|}=-\frac{1}{2}+\frac{1}{2}\log(p(p-1))\\
    &+\sup_{\eta}\left[\Phi(\eta)-\frac{p-2}{4p}\eta^2-\kappa\frac{p-1}{p}m_{\nabla^{2}U} \eta+\frac{\kappa^2}{2p}\left((p-(p-1)\eta^2)V_{\nabla^{2}U}-(p-1)m_{\nabla^{2}U}^2\right)\right]\!+\!O(\kappa^3)\,.\nonumber
\end{align}

Optimization with respect to
$\eta$ has to be carried out separately for the case $p=2$ and $p>2$ due to the vanishing of the $O(1)$ and $\eta^{2}$ term at $p=2$. 

\subsubsection{Optimization for the case of $p>2$}
For this case, the optimal $\eta$ is given by:
\begin{equation}
    \eta^{*}_{(p>2)}\left(\sigma\right) = -\kappa m_{\nabla^{2}U} \frac{2(p-1)}{p-2} + O\left(\kappa^{2}\right) .
\end{equation}
Substitution of $\eta^{*}_{(p>2)}$ into \eqref{eq-25} gives the following: 
\begin{equation}
    \lim_{N\to \infty}\frac{1}{N}\log\overline{\left|\mathrm{det}\nabla^{2}H\left(\sigma\right)\right|} = \frac{1}{2}\left(-1+\log p\left(p-1\right)\right) + \kappa^{2}\left( \frac{1}{2}V_{\nabla^{2}U} + m_{\nabla^{2}U}^{2} \frac{(p-1)}{2(p-2)} \right) + O(\kappa^{3}).
\end{equation}
\subsubsection{Optimization for the case of $p=2$}
For the case of $p=2$, the $-\eta^{2}$ term is absent, leading to a different solution form. Define:
\begin{align}
    \eta_{1}(\sigma) = & \sign\left(-m_{\nabla^{2}U}\right) \left(2 + \frac{1}{4}\kappa^{2}m_{\nabla^{2}U}^{2}\right) + O(\kappa^{3})
    \\
    \eta_{2}(\sigma) = & -\frac{m_{\nabla^{2}U}/\kappa}{V_{\nabla^{2}U}} + O(\kappa)
\end{align}
As \eqref{eq-25} is defined in a piece-wise manner (see definition of $\Phi(\eta)$ above) we take the derivative of \eqref{eq-25} w.r.t $\eta$ and look for extrema in each domain separately:
\begin{equation}
    \frac{d}{d\eta}\lim_{N\to \infty}\frac{1}{N}\log\overline{\left|\mathrm{det}\nabla^{2}H\left(\sigma\right)\right|} =
    \begin{cases}
        -\frac{\kappa}{2}m_{\nabla^{2}U} -\eta\frac{\kappa^2}{2} V_{\nabla^{2}U} + O(\kappa^{3}) & |\eta|<2
        \\
        -\frac{1}{2} \sign\left(\eta\right)\sqrt{\eta ^2-4} - \frac{\kappa}{2}m_{\nabla^{2}U} -\eta\frac{\kappa^2}{2} V_{\nabla^{2}U} + O(\kappa^{3}) & |\eta|>2
    \end{cases}
\end{equation}
We see from the above that a maximum can be found in both domains. Thus, depending on the value of $\eta_{2}$, we have either one or two maxima values s.t. the set of maxima $\eta_{\mathrm{max}}$ is given by:
\begin{equation}
    \eta_{\mathrm{max}} = \begin{cases}
        \left\{\eta_{1},\, \eta_{2}\right\} & |\eta_{2}|<2
        \\
        \eta_{1} & |\eta_{2}|\ge 2
    \end{cases}
\end{equation}
where $|\eta_{1}|>2$ and $\eta_{2}$ is an extremum if it is in the domain $|\eta|<2$. The values of the function at each of the maxima are given by:
\begin{align} \label{eq:log_det_supremuma}
    & \lim_{N\rightarrow\infty}\frac{1}{N}  \log\overline{\left|\mathrm{det}\nabla^{2}H\left(\sigma\right)\right|} \Big |_{\eta=\eta_{1}} = \frac{1}{2}\left(-1+\log 2\right) - \frac{1}{4}\kappa^{2}m_{\nabla^{2}U}^{2} + \kappa^{2}V_{\nabla^{2}U} \left(\left|\eta_{2}\right| - \frac{1}{2}\right)
    \\
    & \lim_{N\rightarrow\infty}\frac{1}{N}  \log\overline{\left|\mathrm{det}\nabla^{2}H\left(\sigma\right)\right|} \Big |_{\eta=\eta_{2}} = \frac{1}{2}\left(-1+\log 2\right) - \frac{1}{4}\kappa^{2}m_{\nabla^{2}U}^{2} + \kappa^{2}V_{\nabla^{2}U} \left(\frac{1}{2} + \frac{1}{4}\eta_{2}^{2}\right)
\end{align} 
Observing the above expressions and noting that $|\eta_{2}| \le 2$ (since $\eta_{2}$ is a solution in the domain $|\eta|<2$) we see that
\begin{equation}
    \lim_{N\rightarrow\infty}\frac{1}{N}  \log\overline{\left|\mathrm{det}\nabla^{2}H\left(\sigma\right)\right|} \Big |_{\eta=\eta_{1}} \le \lim_{N\rightarrow\infty}\frac{1}{N}  \log\overline{\left|\mathrm{det}\nabla^{2}H\left(\sigma\right)\right|} \Big |_{\eta=\eta_{2}} .
\end{equation}
Combining all of the above we find that the optimal $\eta^{*}$ and the supremum of the log-determinant are found to be:
\begin{equation}
    \eta^{*}_{(p=2)}\left(\sigma\right) = 
    \begin{cases} 
        \eta_{2}(\sigma) &\;,\;\;  \left|\eta_{2}\right|<2
        \\
        \eta_{1}(\sigma) \, &\;,\;\;  \left|\eta_{2}\right| \ge 2 
    \end{cases} ,
\end{equation}
\begin{equation} \label{eq:log_det_supremum}
\begin{split}
    & \lim_{N\rightarrow\infty}\frac{1}{N}  \log\overline{\left|\mathrm{det}\nabla^{2}H\left(\sigma\right)\right|} \Big |_{\eta=\eta^{*}}=\\
    &\frac{1}{2}\left(-1+\log 2\right) - \frac{1}{4}\kappa^{2}m_{\nabla^{2}U}^{2} + \kappa^{2}V_{\nabla^{2}U} 
    \begin{cases} 
        \left|\eta_{2}\right| - \frac{1}{2} \, &\;,\;\;  \left|\eta_{2}(\sigma)\right| \ge 2 
        \\
        \frac{1}{2} + \frac{1}{4}\eta_{2}^{2} & \;,\;\; \left|\eta_{2}(\sigma)\right|<2
    \end{cases} .
\end{split}
\end{equation}    
\subsection{Integration over the $N$-sphere} \label{sec:sphere_functional_integration}
Having completed the evaluation of the disorder average $\overline{ \left| \det\left( \nabla^{2}H(\sigma) \right) \right| \delta\left( \nabla H(\sigma) \right) }$, we proceed to evaluate the integration over the surface of the sphere in \eqref{eq:fixed_point_num_def}, i.e.,
\begin{equation}
\begin{split}
    \overline{ \mathcal{N_{\mathrm{tot}}} } = & \left(\frac{1}{2\pi p}\right)^{N/2}\int_{\sqrt{N}\mathbb{S}^{N-1}} d\sigma_{N-1} \exp\left(-\frac{\kappa^{2}}{2 p} \lVert\nabla U\left(\sigma\right)\rVert^{2}  + \log\overline{\left|\mathrm{det}\nabla^{2}H\left(\sigma\right)\right|}\right) .
\end{split}
\end{equation}
The approach we adopt for the evaluation of this integral is to pass from spatial integration over the surface of the sphere to functional integration over the empirical distribution of the coordinates $\sigma_{i}$, which we denote by $\mu$, that is $\mu(x)=N^{-1}\sum_{i=1}^N \delta\left(x-\sigma_i\right)$. First, the integration over the sphere is converted as follows to an integration over $\mathbb{R}^{N}$, such that for any 
smooth function on the sphere,
\begin{equation} \label{eq:sphere_delta_int}
    \int_{\sqrt{N}\mathbb{S}^{N-1}} f(\sigma)d\sigma_{N-1} = \int_{\mathbb{R}^{N}} f(\sigma)\delta\left(\sum_{i=1}^{N}\sigma_{i}^{2} - N\right)d^{N}\sigma.
\end{equation}
This allows us to pass from integration over $\mathbb{R}^{N}$ to functional integration over $\mu$. Noting that for any function $f$ on the sphere,
invariant under permutations of the coordinates, we have $f(\sigma)=f[\mu]$ and therefore applying the Coulomb gas technique \cite{Dyson1962StatisticalI} leads to the following rule for passing between integration on the sphere 
to integration over the space of probability measures on the real line:
\begin{equation}
\begin{split}
    \int_{\sqrt{N}\mathbb{S}^{N-1}} & d\sigma_{N-1}  f\left(\sigma\right) \sim
    \\
    & \frac{S_{N-1}(\sqrt{N})}{\mathcal{Z}}\int D\left[\mu\right] \exp{\left(-N\int dx \mu\log\mu \right)} \: \delta\left(N\int \mu\left(x\right) x^{2}dx - N\right) f\left[\mu\right]
\end{split}
\end{equation}
with $S_{N-1}(\sqrt{N})=\frac{2}{\Gamma(N/2)}(\pi N)^{N/2}$ being the surface area of the $N$-sphere of radius $\sqrt{N}$ and
\begin{equation}
    \mathcal{Z} = \int D\left[\mu\right] \exp{\left(-N\int dx \mu\log\mu \right)} \: \delta\left(N\int \mu\left(x\right) x^{2}dx - N\right).
\end{equation}
Applying this functional integration procedure to the expression for $\overline{ \mathcal{N_{\mathrm{tot}}} }$ results in the following:
\begin{equation} \label{eq:fixed_point_functional_integral}
\begin{split}
    \overline{ \mathcal{N_{\mathrm{tot}}} } = & \frac{C_{N}}{\mathcal{Z}}\int D\left[\mu\right] \delta\left(\int dx \mu\left(x\right) x^{2} - 1\right) \times
    \\
    & \exp{N\left( \frac{1}{N}\log\overline{\left|\nabla^{2}H\left[\mu\right]\right|} -\frac{\kappa^{2}}{2 p N} \lVert\nabla U\left[\mu\right]\rVert^{2} - \int dx \mu\log\mu \right)}
\end{split}
\end{equation}
where $C_{N} = \left(\frac{1}{2\pi p}\right)^{N/2} S_{N-1}(\sqrt{N})$. To derive explicit expressions for $\log\overline{\left|\nabla^{2}H\left[\mu\right]\right|}$ and $\lVert\nabla U\left[\mu\right]\rVert^{2}$, we first calculate the covariant derivatives $\nabla U$ and $\nabla^{2}U$ by exploiting the ``single spin" structure of $U = \sum u\left(\sigma_{m}\right)$:
the covariant derivative and Hessian on the sphere of $U$ are given by:
\begin{equation}
    \left(\nabla U\right)_{m} =  u'\left(\sigma_{m}\right) - \frac{\sigma_{m}}{N}\sum_{l}\sigma_{l}u'\left(\sigma_{l}\right)
\end{equation}
and
\begin{equation}
    \left(\nabla^{2}U\right)_{mn} \approx \delta_{mn}\left( u''\left(\sigma_{m}\right) - \frac{1}{N}\sum_{l}\sigma_{l}u'\left(\sigma_{l}\right)\right)
\end{equation}
where the approximation in the last equality constitutes neglecting terms of rank $O\left(1\right)$ as described in Appendix \ref{Appendix B N-Sphere Covariant Hessian Single Particae} . Using these expressions, we get:
\begin{equation}
    \begin{split}
        \lVert\nabla U\left[\mu\right]\rVert^{2} = \int \mu(x) u'^{2}(x)\, dx-\left(\int x u'(x) \mu(x)  dx\right)^{2} 
    \end{split}
\end{equation}
\begin{equation}
    m_{\nabla^{2}U}\left[\mu\right] = \frac{1}{\sqrt{p(p-1)}}\left(\int u''(x) \mu(x) dx - \int x u'(x) \mu(x)  dx\right)
\end{equation}
\begin{equation} \label{eq:Hessian_var_def}
    V_{\nabla^{2}U}\left[\mu\right] = \frac{1}{p(p-1)}\left(\int u''^{2}(x) \mu(x) dx - \left(\int u''(x) \mu(x)  dx\right)^{2}\right)
\end{equation}
where the explicit form of $\log\overline{\left|\nabla^{2}H\left[\mu\right]\right|}$ is found by replacing $m_{\nabla^{2}U}\left(\sigma\right) \to m_{\nabla^{2}U}\left[\mu\right]$ and $V_{\nabla^{2}U}\left(\sigma\right) \to V_{\nabla^{2}U}\left[\mu\right]$.
Since the integral in the numerator of $\overline{\mathcal{N_{\mathrm{tot}}} }$ is of the form $\int D[\mu] \exp{N S[\mu]}$, it is evaluated by the saddle point technique in the large $N$ limit. We start by writing down the functional derivative of the exponent w.r.t $\mu$:
\begin{equation} \label{eq:spatial_saddle_point}
    \begin{split}
    \frac{\delta S}{\delta \mu} = & \alpha +\beta x^2 -\log\mu - 1 +\lim_{N\to \infty} \frac{1}{N}\frac{\delta\log\overline{\left|\mathrm{det}\nabla^{2}H\left[\mu\right]\right|}}{\delta \mu\left(x\right)}
    \\
    &  -\kappa^{2}\frac{1}{2p}\left(u'^{2}(x)-2\left(\int y u'(y) \mu(y)  dy\right)x u'(x)\right) = 0
    \end{split}
\end{equation}
where $\alpha$ and $\beta$ are Lagrange multipliers associated with the constraints on the normalization and second moment of $\mu$.  We solve the above equation for $\mu$ and the Lagrange multipliers $\alpha, \, \beta$ by expanding each in a perturbative expansion as follows:
\begin{equation} \label{eq:perturbative_expansion_def}
    \begin{split}
        & \mu\left(x\right) = \mu_{0}\left(x\right) + \kappa\mu_{1}\left(x\right) +  \kappa^{2}\mu_{2}\left(x\right) + O\left(\kappa^{3}\right)
        \\
        & \beta = \beta_{0} + \kappa\beta_{1} + \kappa^{2}\beta_{2} + O\left(\kappa^{3}\right)
        \\
        & \alpha = \alpha_{0} + \kappa\alpha_{1} + \kappa^{2}\alpha_{2} + O\left(\kappa^{3}\right).
    \end{split}
\end{equation}
As the expression for $\log\overline{\left|\mathrm{det}\nabla^{2}H\left[\mu\right]\right|}$ is different for the cases $p>2$ and $p=2$ (which includes two cases of its own) we solve the saddle point equations for these cases separately.
\subsubsection{The case of $p=2$}
Since for the case of $p=2$ the log-determinant is defined in a piecewise form, we evaluate the integral over the measure $\mu$ separately for each case as well as for the boundary between the cases i.e.
\begin{equation}
    \overline{\mathcal{N_{\mathrm{tot}}}} = \int\displaylimits_{\mathcal{M}} D[\mu] \exp{N S[\mu]} = \int\displaylimits_{\mathcal{M}_{<}} D[\mu] \exp{N S_{<}[\mu]} + \int\displaylimits_{\mathcal{M}_{\ge} } D[\mu] \exp{N S_{\ge}[\mu]}
\end{equation}
where $\mathcal{M}$ is the space of measures with a unit second moment, $\mathcal{M}_{<}$, $\mathcal{M}_{\ge}$ denote the sets $\left\{\mu\in\mathcal{M}:\eta_{2}[\mu]<2\right\}$, $\left\{\mu\in\mathcal{M}:\eta_{2}[\mu]\ge2\right\}$, respectively, and $S_{<}[\mu]$, $S_{\ge}[\mu]$ denote the exponent in the cases $\eta_{2}<2$ and $\eta_{2} \ge 2$, respectively. Next, we evaluate each of the integrals by the saddle point method. Thus, $\Sigma_{\mathrm{tot}}$ is given by:
\begin{equation}
    \Sigma_{\mathrm{tot}} = \max \left\{\sup\displaylimits_{\mu\in \mathcal{M}_{<}} S_{<}[\mu] \quad ,\sup\displaylimits_{\mu\in \mathcal{M}_{\ge} } S_{\ge}[\mu] \right\}
\end{equation}
In addition, we evaluate the integral on the boundary between the two branches to make sure that the global maximum is indeed found. 
\paragraph{The case of $\left|\eta_{2}\right| \ge 2$}
For this case, the absolute value of the determinant of the Hessian  and its functional derivative are given by:
\begin{align}
    \lim_{N\to \infty}\frac{1}{N}\log\overline{\left|\mathrm{det}\nabla^{2}H\left[\mu\right]\right|}&= \frac{1}{2}\left(-1+\log 2\right) + \kappa m_{\nabla^{2}U}\left[\mu\right] - \frac{\kappa^{2}}{4}\left(m_{\nabla^{2}U}^{2}\left[\mu\right]+2V_{\nabla^{2}U}\left[\mu\right]\right) + O\left(\kappa^{3}\right)\\
    \lim_{N\to \infty}\frac{1}{N}\frac{\delta\log\overline{\left|\mathrm{det}\nabla^{2}H\left[\mu\right]\right|}}{\delta \mu\left(x\right)}&= 
      \frac{\kappa}{\sqrt{2}}\left(u''(x)-x u'(x)\right) -  \frac{\kappa^{2}}{4}\left(u''(x)^2+2u''(x) \int \mu_{0}(y) u''(y) \, dy\right) + O(\kappa^{3}).
\end{align}
Substitution of this expression, along with the perturbative expansions given above in the saddle point condition in eq. \eqref{eq:spatial_saddle_point} gives the following equations:
\begin{align}
         O(\kappa^{0}):\,& \alpha_{0}+\beta_{0}x^2-\log (\mu_{0}(x)) = 0
        \\
         O(\kappa^{1}):\,& \alpha_{1}+\beta_{1}x^2-\frac{\mu_{1}(x)}{\mu_{0}(x)} +\frac{u''(x)-x u'(x)}{\sqrt{2}}=0
        \\
         O(\kappa^{2}):\, &\alpha_{2}+\beta_{2}x^2-\frac{ \mu_{2}(x)}{\mu_{0}(x)}+\frac{\mu_{1}(x)^2}{2\mu_{0}(x)^2} -\frac{ u''(x)^2+u'(x)^2}{4}-\frac{u''(x)-x u'(x)}{2}\int \mu_{0}(y) u''(y) \, dy = 0.
\end{align}
Note that we have used the identity for the law $\mu_{0}(x) = \exp(-x^{2}/2)/\sqrt{2\pi}$:
\begin{equation}
    \int \mu_{0}(y) u''(y) \, dy = \int \mu_{0}(y) y u'(y) \, dy,
\end{equation}
and that the Lagrange multiplier expansion terms $\alpha_{l},\, \beta_{l}$ are determined by the constraints:
\begin{equation} \label{eq:mu_constraints}
    \begin{split}
        & \int \mu_{l}\left(x\right) dx = \delta_{l0}
        \\
        & \int x^{2} \mu_{l}\left(x\right) dx = \delta_{l0}.
    \end{split}
\end{equation}
Solving these equations gives the following expressions for $\mu$:
\begin{align}
    \mu_{0}(x) =& \frac{1}{\sqrt{2\pi}}e^{-x^{2}/2}
        \\
    \mu_{1}(x) =& \frac{\mu_{0}(x)}{\sqrt{2}}\left(\frac{1-x^2}{2}\int  \mu_{0}(y) y^2 \left(u''(y)-y u'(y)\right)dy+u''(x)-x u'(x)\right)
        \\
    \mu_{2}(x) =& \mu_{0}(x)\left(\alpha_{2} x^2+\beta_{2}-\frac{u'(x)^2+u''(x)^2}{4} -\frac{u''(x)-x u'(x)}{2} \left(\int \mu_{0}(y) u''(y) \, dy\right)\right)-\frac{\mu_{1}(x)^2}{2\mu_{0}(x)}\,,
\end{align}
where $\alpha_2$ and $\beta_2$ have a cumbersome expression, obtained self-consistently by imposing that  \eqref{eq:mu_constraints} hold. Using this expression, we can evaluate the average log-determinant: 
\begin{align}
    &\lim_{N\to \infty}\frac{1}{N}\log\overline{\left|\mathrm{det}\nabla^{2}H\left[\mu\right]\right|}=\\
    &\frac{1}{2}\left(-1+\log 2\right)+\frac{\kappa^2}{\sqrt{2}}\int dx\,\mu_1(x)\,(u''(x)-x u'(x))-\frac{\kappa^2}{4}\left[\int dx\,\mu_0(x)\,u''(x)^2-\left(\int dx\,\mu_0(x)\,u''(x)\right)^2\right]+O(\kappa^3)\nonumber
\end{align}

Next, we can substitute the expressions we found for the log-determinant and $\mu$ in the expression for the
annealed complexity to get
\begin{align}
    \Sigma_{\eta \ge 2}^{(p=2)} = & \lim_{N\rightarrow\infty} \frac{1}{N} \log \overline{ \mathcal{N}_{\mathrm{tot}} }\\
    =&\lim_{N\to \infty}\frac{1}{N}\log C_N-\lim_{N\to \infty}\frac{1}{N}\log {\cal Z}+\lim_{N\to \infty}\frac{1}{N}\log\overline{\left|\mathrm{det}\nabla^{2}H\left[\mu\right]\right|} - \int dx \mu\log\mu -\frac{\kappa^{2}}{4} \lVert\nabla U\left[\mu\right]\rVert^{2}\;.\nonumber
\end{align}
Let us first use that in the large $N$ limit
\begin{equation}
    \lim_{N\to \infty}\frac{1}{N}\log C_N=\lim_{N\to \infty}\frac{1}{N}\log S_{N-1}-\frac{1}{2}\log(2\pi p)=\frac{1}{2}\left(1-\log(p)\right)\;.
\end{equation}
We may then evaluate the quantity involving ${\cal Z}$ by the Laplace method, as
\begin{align}
    &\lim_{N\to \infty}\frac{1}{N}\log {\cal Z}=\max_{\mu,\alpha,\beta}\left[-\int dx\, \mu(x)\log \mu(x)+\alpha\left(\int dx\,\mu(x)-1\right)+\beta\left(\int dx\,x^2\,\mu(x)-1\right) \right]\;.
\end{align}
One can easily check that the maximum is reached for $\mu(x)=\mu_0(x)$ and thus
\begin{equation}
    \lim_{N\to \infty}\frac{1}{N}\log {\cal Z}=-\int dx\, \mu_0(x)\log \mu_0(x)\;.
\end{equation}
The entropic term can simply be developped up to second order in $\kappa$, yielding
\begin{align*}
    -\int dx\,\mu(x)\ln \mu(x)=& -\int dx\,\mu_0(x)\ln \mu_0(x)-\kappa \int dx \mu_1(x)\left(1-\frac{1}{2}\ln(2\pi)-\frac{x^2}{2}\right)\\
    &-\kappa^2\int dx \left[\mu_2(x)\left(1-\frac{1}{2}\ln(2\pi)-\frac{x^2}{2}\right)+\frac{\mu_1(x)^2}{2\mu_0(x)}\right]+O(\kappa^3)\\
    =& -\int dx\,\mu_0(x)\ln \mu_0(x)-\frac{\kappa^2}{2}\int dx\,\frac{\mu_1(x)^2}{\mu_0(x)}+O(\kappa^3)\\
    =&-\int dx\,\mu_0(x)\ln \mu_0(x)-\frac{\kappa^2}{2\sqrt{2}}\int dx\,\mu_1(x)(u''(x)-x u'(x))+O(\kappa^3)\;,\nonumber
\end{align*}
where we have used the explicit expression of $\mu_0(x)$ as well as  \eqref{eq:mu_constraints}. Finally, as the term $-\kappa^2\lVert\nabla U\left[\mu\right]\rVert^{2}/4$ is already of order $\kappa^2$, it can directly be evaluated for $\mu_0$ for the purpose of our perturbative computation and reads
\begin{equation}
    -\frac{\kappa^2}{4}\lVert\nabla U\left[\mu\right]\rVert^{2}=-\frac{\kappa^2}{4}\left[\int dx\,\mu_0(x)\,u'(x)^2-\left(\int dx\,\mu_0(x)\,x u'(x)\right)^2\right]+O(\kappa^3)
\end{equation}
Gathering all terms and some straightforward algebra lead to the following complexity:
\begin{equation}
    \begin{split}
        \Sigma_{\eta \ge 2}^{(p=2)} = & \lim_{N\rightarrow\infty} \frac{1}{N} \log \overline{ \mathcal{N}_{\mathrm{tot}} } = 
        \\
        & \frac{\kappa^{2}}{2}\left(- \frac{1}{4} \left(\int x^2 \mu_{0}(x) \left(u''(x)-x u'(x)\right) \, dx\right)^{2} + \left(\int dx \mu_{0} \left(x\right) u''(x)\right)^{2}  \right.
        \\
        & \left. +  \frac{1}{2}\int dx \left(x^2-1\right) \mu_{0}(x) u'(x)^2 
        - \int dx \, x \mu_{0}(x) u'(x) u''(x) \right) + O(\kappa^{3})\;.
    \end{split}
\end{equation}      
By using the identity $\overline{ f'(x) }_{\mu_{0}} = \overline{ x f(x) }_{\mu_{0}}$ (as $\mu_{0}$ is the Gaussian) we get:
\begin{align} \label{eq:Gaussian_integral_identities}
        - \frac{1}{4} \left(\int x^2 \mu_{0}(x) \left(u''(x)-x u'(x)\right) \, dx\right)^{2} + \left(\int dx \mu_{0} \left(x\right) u''(x)\right)^{2} = &  0
        \\
        \frac{1}{2}\int dx \left(x^2-1\right) \mu_{0}(x) u'(x)^2 
        - \int dx \, x \mu_{0}(x) u'(x) u''(x) = & 0
\end{align}
thus we are left with
\begin{equation}
    \Sigma_{\eta \ge 2}^{(p=2)} = 0 + O(\kappa^{3}).
\end{equation}

Next, to check the nature of the extremum, we calculate the second functional derivative of $S[\mu]$ evaluated at $\mu^{*}$:
\begin{equation}
    \frac{\delta^{2}S[\mu]}{\delta \mu(x) \delta \mu(y)}\Bigg\vert_{\mu^{*}} = - \delta\left(x-y\right)\left(\frac{1}{\mu_{0}(x)}+O(\kappa)\right) + \frac{\kappa^{2}}{2}\left( u''(x) u''(y) + x u'(x) y u'(y) \right) + O(\kappa^{3})
\end{equation}
We consider the RHS as an operator in $L^2(\mu_0)$. By assumption, both $u''$ and $xu'$ are bounded in $L^2(\mu_0)$, and using that $\kappa^2$ is small, we deduce that this operator is negative definite. In particular,
the extremum we found is a (local) maximum.

Finally, we derive conditions for self-consistency of this solution by requiring that $\left|\eta_{2}[\mu^{*}]\right| > 2$. Substitution of $\mu^{*}$ in $\eta_{2}[\mu^{*}]$ gives:
\begin{equation}
\begin{split}
    \eta_{2}[\mu^{*}] = & \frac{m_{\nabla^{2}U}[\mu^{*}]}{\kappa V_{\nabla^{2}U}[\mu^{*}]} =  \frac{m_{\nabla^{2}U}[\mu_{1}]}{V_{\nabla^{2}U}[\mu_{0}]} + O(\kappa) 
    \\
    = & \frac{\frac{1}{2}\left(\int x^2 \mu_{0}(x) \left(u''(x)-x u'(x)\right) \, dx\right)^{2} + \int\mu_{0}(x) \left(u''(x)-x u'(x)\right)^{2} \, dx}{\int\mu_{0}(x) u''(x)^{2} \, dx - \left(\int\mu_{0}(x) u''(x) \, dx\right)^{2}}
\end{split}
\end{equation}
The requirement $\eta_{2}[\mu^{*}]>2$ gives after some algebra and use of \eqref{eq:Gaussian_integral_identities} (we select $m_{\nabla^{2}U}>0$ without loss of generality):
\begin{equation}
    \eta_{2}[\mu^{*}] > 2 \iff \int \mu_{0}(x)\left(u''(x)^{2} - u'(x)^{2}\right) \, dx < 0
\end{equation}
Thus this solution is self-consistent when the other ($\eta_{2}<2$) branch has negative complexity i.e. $\Theta^{(2)}[u]<0$,
see \eqref{eq-Thetag}. This is consistent with the picture of the $\eta_{2}>2$ branch corresponding to zero added complexity at leading order.
\paragraph{The case of $\left|\eta_{2}\right|<2$}

For this case,  the absolute value of the determinant of the Hessian  and its functional derivative are given by:
\begin{align}
    \lim_{N\to \infty}\frac{1}{N}\log\overline{\left|\mathrm{det}\nabla^{2}H\left[\mu\right]\right|} = &\frac{1}{2}\left(-1+\log 2\right)+ \frac{1}{4}\kappa^{2}\left(2V_{\nabla^{2}U}\left[\mu\right]+\frac{m_{\nabla^{2}U}\left[\mu\right]^{2}}{\kappa^{2}V_{\nabla^{2}U}\left[\mu\right]}+m_{\nabla^{2}U}\left[\mu\right]^{2}\right) + O\left(\kappa^{3}\right).
\end{align}
Then,
\begin{equation}
\begin{split}
    \lim_{N\to \infty}\frac{1}{N}\frac{\delta\log\overline{\left|\mathrm{det}\nabla^{2}H\left[\mu\right]\right|}}{\delta \mu\left(x\right)} = &\kappa  \frac{ m_{\nabla^{2}U}[\mu_{1}]}{2V_{\nabla^{2}U}\left[\mu_{0}\right]}\left(u''(x)-x u'(x)\right) +
    \\
    &   \frac{\kappa^{2}}{2}  \left(\frac{ m_{\nabla^{2}U}[\mu_{2}]}{ V_{\nabla^{2}U}\left[\mu_{0}\right]}-\frac{ m_{\nabla^{2}U}[\mu_{1}]  V_{\nabla^{2}U}\left[\mu_{1}\right]}{\left( V_{\nabla^{2}U}\left[\mu_{0}\right]\right)^{2}}\right) \left(u''(x)-x u'(x)\right)\\
    &-\kappa^2 \frac{ m_{\nabla^{2}U}[\mu_{1}]^2 \left(u''(x)^2-2 u''(x) \int \mu_{0}(y) u''(y) \, dy\right)}{4 \left( V_{\nabla^{2}U}\left[\mu_{0}\right]\right)^{2}}
    \\
    & + \frac{\kappa^{2}}{2}\left(u''(x)^2-2 u''(x) \left(\int \mu_{0}(y) u''(y) \, dy\right)\right)+ O(\kappa^{3})+o_N(1).
\end{split}
\end{equation}
Substitution of this expression, along with the perturbative expansions for $\mu$ in  \eqref{eq:spatial_saddle_point} gives the following equations:
\begin{equation}
    \begin{split}
        O(\kappa^{0}):&\, \alpha_{0}+\beta_{0}x^2-\log (\mu_{0}(x)) = 0
        \\
        O(\kappa^{1}):& \, \alpha_{1}+\beta_{1}x^2-\frac{\mu_{1}(x)}{\mu_{0}(x)} + \frac{ m_{\nabla^{2}U}[\mu_{1}]}{2 V_{\nabla^{2}U}\left[\mu_{0}\right]}\left(u''(x)-x u'(x)\right)=0
        \\
        O(\kappa^{2}):& \, \alpha_{2}+\beta_{2}x^2-\frac{ \mu_{2}(x)}{\mu_{0}(x)} + \left(\frac{ m_{\nabla^{2}U}\left[\mu_{2}\right] }{2 V_{\nabla^{2}U}\left[\mu_{0}\right]} - \int \mu_{0}(y) u''(y) \, dy\right)\left(u''(x)-x u'(x)\right)+\frac{1}{2}\left(u''(x)^2-u'(x)^2\right) = 0
    \end{split}
\end{equation}
where we have used that all first order correction terms vanish in the second order equation, as can easily be deduced from the first order equation. The Lagrange multiplier expansion terms $\alpha_{l},\, \beta_{l}$ are determined by \eqref{eq:mu_constraints}
as in the calculation of the previous case.
Solving these equations gives the following:
\begin{equation*}
    \begin{split}
        & \mu_{0}(x) = \frac{1}{\sqrt{2\pi}}e^{-x^{2}/2},
        \\
        & \mu_{1}(x) = 0,
        \\
        & \mu_{2}(x) = \mu_0(x)\left[\alpha_{2} x^2+\beta_{2}+\frac{1}{2}\left(u''(x)^2-u'(x)^2\right)+\left(u''(x)-x u'(x)\right) \left(\frac{ m_{\nabla^{2}U}\left[\mu_{2}\right] }{2 V_{\nabla^{2}U}\left[\mu_{0}\right]} - \int \mu_{0}(y) u''(y) \, dy\right)\right].
    \end{split}
\end{equation*}
Next, we can substitute the expression for $\mu$ in the expression for the number of critical points, use that $\mu_1=0$ and use the expansions derived above to find
\begin{equation}
\begin{split}
    \overline{ \mathcal{N}_{\mathrm{tot}} } = &
    C_{N}\exp{N\left( \frac{1}{N}\log\overline{\left|\mathrm{det}\nabla^{2}H\left[\mu\right]\right|} -\frac{\kappa^{2}}{4} \lVert\nabla U\left[\mu_{0}\right]\rVert^{2}+O(\kappa^{3})+o_N(1)\right)}
    \\
    = & C_{N}\exp{N\left(\frac{1}{2}\left(-1+\log 2\right)+\kappa^{2}\frac{1}{2} V_{\nabla^{2}U}\left[\mu_{0}\right]-\frac{\kappa^{2}}{4} \lVert\nabla U\left[\mu_{0}\right]\rVert^{2}+O(\kappa^{3})+o_N(1)\right)}.
\end{split}
\end{equation}
Gathering all terms and some straightforward algebra lead to the following complexity:
\begin{equation}
    \begin{split}
        \Sigma_{\eta_{2} < 2}^{(p=2)} = \lim_{N\rightarrow\infty} \frac{1}{N} \log \overline{ \mathcal{N}_{\mathrm{tot}} } = & \kappa^{2} \frac{1}{4}\int dx \mu_{0}(x)\left(u''(x)^2-u'(x)^2\right) + O(\kappa^{3}) 
        = \kappa^{2} \Theta^{(2)}\left[u(x)\right] + O(\kappa^{3})+o_N(1)
    \end{split}
\end{equation}   
To verify the type of extremum we found in this case, we proceed to evaluate the contribution of the integral on the boundary $\eta_{2}=2$ in what follows.

\paragraph{The case of $\left|\eta_{2}\right| = 2$}
Since $\log \det \left|\nabla^{2}H\right|$ and $\partial_{\eta_{2}}\log \det \left|\nabla^{2}H\right| $ are continuous w.r.t to $\eta_{2}$ at $\eta_{2}=2$ (see  \eqref{eq:log_det_supremum}), we can pick either one of the branches for the following calculation. We pick the expressions for the case $\eta_{2}>2$:
\begin{equation}
\begin{split}
    \lim_{N\to \infty}\frac{1}{N}\log\overline{\left|\mathrm{det}\nabla^{2}H\left[\mu\right]\right|} = & \frac{1}{2}\left(-1+\log 2\right) + \kappa m_{\nabla^{2}U}\left[\mu\right] + \frac{1}{4}\kappa^{2}\left(m_{\nabla^{2}U}\left[\mu\right]-2V_{\nabla^{2}U}\left[\mu\right]\right) 
    \\
    + & \gamma\left(m_{\nabla^{2}U} - \kappa 2  V_{\nabla^{2}U} \right) +  O\left(\kappa^{3}+o_N(1)\right)\,.
\end{split}
\end{equation}
where we have introduced an additional Lagrange multiplier $\gamma$ to enforce the constraint $\eta_{2}[\mu]=2$. Taking the functional derivative of $S[\mu]$ under this constraint and consecutive substitution of perturbative expansions give the following equations for the zero and first order in $\kappa$:
\begin{equation}
    \begin{split}
        & O(\kappa^{0}):\, \alpha_{0}+\beta_{0}x^2 + \gamma_{0}\frac{1}{\sqrt{2}}\left( u''(x)-x u'(x)\right)-\log (\mu_{0}(x)) = 0
        \\
        & O(\kappa^{1}):\, \alpha_{1}+\beta_{1}x^2-\frac{\mu_{1}(x)}{\mu_{0}(x)} \!+\!\Big(\frac{1}{\sqrt{2}}+\gamma_{1}\Big)\left( u''(x)-x u'(x)\right) + \gamma_{0}\sqrt{2}\left(-u''(x)^{2}+u''(x)\overline{u''}\right)=0\,.
    \end{split}
\end{equation}
where here the $\overline{f(x)}$ notation denotes averaging w.r.t the standard Gaussian measure $\mu_{0}$. Solving these equations, under the three constraints on $\mu$ (total mass equal to one, unit second moment and $\eta_{2}[\mu]=2$) gives the following solutions:
\begin{equation}
    \begin{split}
        & \mu_{0}(x) = \frac{1}{\sqrt{2\pi}}e^{-x^{2}/2}
        \\
        & \mu_{1}(x) =\frac{\mu_0(x)}{\sqrt{2}}\left(1 + \frac{\Theta^{(2)}[u(x)]}{m_{\nabla^{2}U}[\mu_{1,\eta_{2}>2}]}\right)\left(\frac{1}{2}\left(x^2-1\right)\int  \mu_{0}(y) y^2 \left(u''(y)-y u'(y)\right)dy+u''(x)-x u'(x)\right)\,.
    \end{split}
\end{equation}
where we found that $\gamma = \kappa \frac{\Theta^{(2)}[u(x)]}{\sqrt{2}m_{\nabla^{2}U}[\mu_{1,\eta_{2}>2}]} + O(\kappa^{2})$. and denoted by $\mu_{1,\eta_{2}>2}$ the first order correction to $\mu^{*}$ in the case $\eta_{2}>2$.
Substitution of these solutions in $S[\mu]$ results in the following complexity:
\begin{equation}
    \Sigma_{\eta_{2}=2} = \kappa^{2}\left(\Theta^{(2)}[u]\left(1 - \frac{c}{\sqrt{2}}\right) - \frac{c}{4}\left(\Theta^{(2)}[u]\right)^{2}\right) + O(\kappa^{3})
\end{equation}
where we have denoted by $c>0$ the following expression:
\begin{equation}
    c = 1 + \frac{\left(\overline{y^{2}\left(u''(y) - y u'(y)\right)}_{\mu_{0}}\right)^{2}}{\frac{1}{2}\left(\overline{y^{2}\left(u''(y) - y u'(y)\right)}_{\mu_{0}}\right)^{2} + \overline{\left(u''(y) - y u'(y)\right)^{2}}_{\mu_{0}}}\,.
\end{equation}
Now we are in a position to compare the complexity resulting from the boundary $\eta_{2}=2$ with the complexity of the extrema found for the case $\eta_{2}<2$. As $c>0$ we can see that in the case where $\Theta^{(2)}>0$ we have that $\Sigma_{\eta_{2} < 2} > \Sigma_{\eta_{2}=2}$. Thus, we conclude that the extermum found for the case $\eta_{2} < 2$ is a maximum.
\subsubsection{The case of $p>2$}
For this case, the log determinant is given by:
\begin{align}
    &\lim_{N\to\infty}\frac{1}{N}\log\overline{\left|\mathrm{det}\nabla^{2}H\left[\mu\right]\right|} =\\
    &\frac{1}{2}\log p\left(p-1\right) + \kappa^{2}\left( \frac{1}{2}V_{\nabla^{2}U}\left[\mu\right] + m_{\nabla^{2}U}\left[\mu\right]^{2} \frac{(p-1)^{2}}{4p(p-2)} \right) + O(\kappa^{3}).\nonumber
\end{align}
It is clear from the previous cases that $m_{\nabla^{2}U} = O\left(\kappa\right)$ and thus, to leading order in $\kappa$, the complexity is given by:
\begin{equation}
\begin{split}
    \Sigma^{(p>2)} & =  \frac{1}{2}\log\left(p-1\right) + \kappa^{2}\left(\frac{1}{2}V_{\nabla^{2}U}\left[\mu_{0}\right] - \frac{1}{2 p} \lVert\nabla U\left[\mu_{0}\right]\rVert^{2}\right) + O\left(\kappa^{3}\right) 
    \\
    & 
    = \frac{1}{2}\log\left(p-1\right) + \kappa^{2}\Theta^{(p)}\left[u(x)\right] + O\left(k^{3}\right).
\end{split}
\end{equation}

\newpage

\section{Evaluation of the Mean Number of Critical Points of a Given Index}
Here we aim to evaluate the number of critical points of a given index $k$ (number of negative eigenvalues of the Hessian). We recall \eqref{eq:fixed_point_k_num_def}.
Taking identical steps as for the case of $\overline{ N_{\mathrm{tot}} }$, we arrive at the average value $\overline{ \left| \det\left( \nabla^{2}H \right) \right|\delta\left( I\left(\nabla^{2} H\right) - k \right)}$. This is evaluated as follows:
\begin{equation}
\begin{split}
     & \overline{ \left| \det\left( \nabla^{2}H \right) \right|  \delta\left( I\left(\nabla^{2} H\right) - k \right)} =  \overline{ \left| \det\left( \nabla^{2} \phi + \kappa\nabla^{2}U\right) \right|\delta\left( I\left(\nabla^{2} \phi + \kappa\nabla^{2}U\right) - k \right) } 
     \\
     = & \int_{\mathbb{R}}\frac{\exp{\left(-\frac{Ns^{2}}{2p^2}\right)}}{\sqrt{2\pi p^{2}/N}}\overline{ \left|\mathrm{det} \tilde{M} + \kappa\nabla^{2}U + s\mathrm{I}_{N-1}\right|\delta\left( I\left(\tilde{M} + \kappa\nabla^{2}U + s\mathrm{I}_{N-1} \right) - k \right) }
\end{split}
\end{equation}
with $\tilde{M}\sim \sqrt{p(p-1)\frac{N-1}{N}}\mathrm{GOE}_{N-1}$. We first evaluate the GOE averaging as in \cite{Arous2021LandscapeManifold} and follow by the integration over $\mathbb{R}$ to get:
\begin{align*} 
    &\overline{\left|\mathrm{det} \nabla^{2}H\left(\sigma\right)\right|\delta\left( I\left(\nabla^{2} H\right) - k \right)} \\
    &= \exp{N \left(\int dx \, \nu_{\nabla^{2}H} \left(x\right) \log\left|x-s^{*}\left(a\right)\right| - \frac{s^{*}\left(a\right)^{2}}{2 p^{2}}+o_N(1)\right)},
\end{align*}
where we have explicitly assumed that $k$ is extensive i.e. the ratio $a = k/N$ is fixed when the large $N$ limit is taken. Also, $s^{*}\left(a\right)$ is determined by the index constraint:
\begin{equation}
    \int_{-\infty}^{0}\nu_{\nabla^{2}H} \left(x + s^{*}\left(a\right)\right) dx = a.
\end{equation}
and $\nu_{\nabla^{2}H}$ is the widened and shifted semicircular distribution found in \eqref{eq:hessian_spectral_dist}. Note that the above expression is very similar in form to the one considered in the case of the total complexity with the difference that here $s$ is set by the index whereas in the previous case the complexity is the supremum over $s$. Thus, we can utilize the same expressions derived in the previous section to give:
\begin{equation} \label{eq:log_det_k}
\begin{split}
    &\lim_{N\to \infty} \frac{1}{N}\log\overline{\left|\mathrm{det}\nabla^{2}H\left(\sigma\right)\right|\delta\left( I\left(\nabla^{2} H\right) - k \right)} = \Phi\left(\eta\right) + \frac{1}{2}\log p\left(p-1\right)
    \\
    & -\frac{p-2}{4p}\eta^{2} + \kappa\frac{p-1}{p} m_{\nabla^{2}U}\eta +  \kappa^{2}\left(\frac{V_{\nabla^{2}U}}{2}\left(1 - \frac{(p-1)\eta^{2}}{p}\right)-m_{\nabla^{2}U}^{2}\frac{p-1}{2p}\right) + O\left(\kappa^{3}\right)
\end{split}
\end{equation}
where $\eta$ is as in the calculation for the total complexity, see \eqref{eq-eta}. Thus, the choice of index $k = a N$ fixes the parameter $\eta$. 

As in the case of the total complexity, we proceed to carry out the integration over the $N$-sphere by passing to functional integration over the measure $\mu$. The functional derivative of $\frac{1}{N}\log\overline{\left|\mathrm{det}\nabla^{2}H\left[\mu\right]\right|}$ is given by:
\begin{equation}
\begin{split}
    & \lim_{N\to \infty}\frac{1}{N}\frac{\delta\log\overline{\left|\mathrm{det}\nabla^{2}H\left[\mu\right]\right|}}{\delta \mu\left(x\right)} = \kappa  \sqrt{\frac{p-1}{p^{3}}} \eta \left(u''(x)-x u'(x)\right)
    \\
    &  +  \frac{\kappa^{2}}{p(p-1)}\left(-\frac{1}{2}\left(\frac{p-1}{p}\eta^{2}-1\right) u''(x)^2+u''(x)\left(\frac{p-1}{p}\eta^{2}-1\right) \int \mu_{0}(y) u''(y) \, dy\right) + O(\kappa^{3}).
\end{split}
\end{equation}
Substitution of this expression, along with the perturbative expansions given above in eq. \eqref{eq:spatial_saddle_point} gives the following equations:
\begin{equation}
    \begin{split}
        & O(\kappa^{0}):\, \alpha_{0}+\beta_{0}x^2-\log (\mu_{0}(x)) = 0
        \\
        & O(\kappa^{1}):\, \alpha_{1}+\beta_{1}x^2-\frac{\mu_{1}(x)}{\mu_{0}(x)} + \sqrt{\frac{p-1}{p^{3}}}\eta\left(u''(x)-x u'(x)\right)=0.
    \end{split}
\end{equation}
Solving these equations, along with the constraints as above, gives the following solutions:
\begin{equation} \label{eq:mu_k}
    \begin{split}
        & \mu_{0}(x) = \frac{1}{\sqrt{2\pi}}e^{-x^{2}/2}
        \\
        & \mu_{1}(x) =  \mu_{0}(x)\,\sqrt{\frac{p-1}{p^{3}}}\eta\left(\frac{1}{2}\left(1-x^2\right)\int  \mu_{0}(y) y^2 dy \left(u''(y)-y u'(y)\right)+u''(x)-x u'(x)\right).
    \end{split}
\end{equation}
Here we focus only on the solutions up to $O(\kappa)$ since $\mu_{2}$ does not affect the complexity to leading order as found for the total complexity.
Substitution of these results in the expression for the log-determinant leads to the following expression for the number of critical points:
\begin{equation} \label{eq:num_points_k_pert_expansion}
\begin{split}
    \frac{1}{N}\log\frac{\overline{ \mathcal{N}_{a} }}{C_{N}}= &
   \left( \frac{1}{N}\log\overline{\left|\mathrm{det}\nabla^{2}H\left[\mu\right]\right|} -\kappa^{2}\frac{1}{2}\int dx \frac{\mu_{1}^{2}}{\mu_{0}} -\frac{\kappa^{2}}{2 p} \lVert\nabla U\left[\mu_{0}\right]\rVert^{2}+O(\kappa^{3})\right)
    \\
    = & \Phi\left(\eta\right) + \frac{1}{2}\log p\left(p-1\right) -\frac{p-2}{4p}\eta^{2}
    \\
    &  +\kappa^{2}\left(\frac{p-1}{p}\eta m_{\nabla^{2}U}[\mu_1] +\frac{1}{2}\left(1 - \frac{\eta^{2}(p-1)}{p}\right) V_{\nabla^{2}U}\left[\mu_{0}\right] -\frac{1}{2}\int dx \frac{\mu_{1}^{2}}{\mu_{0}} -\frac{1}{2 p} \lVert\nabla U\left[\mu_{0}\right]\rVert^{2}\right)
    +O(\kappa^{3}),
\end{split}
\end{equation}
where as before $C_N=(\frac{1}{2\pi p})^{N/2}S_{N-1}(\sqrt{N})$. Next, we observe that $\mu_{1,k} = \sqrt{\frac{p-1}{p^{3}}} \eta \mu_{1,\eta>2}$ (where $\mu_{1,\eta>2}$ denotes the first order term $\mu_{1}$ for the case $\eta_{2}>2$ and $p=2$) and make this substitution to get:
\begin{equation}
\begin{split}
    \Sigma_a^{(p)}[u]=&\lim_{N\to\infty}\frac{1}{N}\ln\overline{ \mathcal{N}_{a} }=\Phi\left(\eta\right) + \frac{1}{2}\log \left(p-1\right) -\frac{p-2}{4p}\eta^{2} +  
    \\
    &
    \kappa^{2}\left(\frac{2(p-1)}{p^3}\eta^{2} m_{\nabla^{2}U}[\mu^{(2)}_{1,\eta>2}] +\frac{1}{p(p-1)}\left(1 - \frac{\eta^{2}(p-1)}{p}\right) V^{(2)}_{\nabla^{2}U}\left[\mu_{0}\right]  \right.
    \\
    & 
    \left. -\frac{p-1}{p^{3}}\eta^{2}\int dx \frac{(\mu^{(2)}_{1,\eta>2})^{2}}{\mu_{0}} -\frac{1}{2 p} \lVert\nabla U\left[\mu_{0}\right]\rVert^{2}\right)+O(\kappa^{3}).
\end{split}
\end{equation}
Simple algebraic manipulations give the following expression:
\begin{equation} \label{eq:complexity_k_result}
\begin{split}
    \Sigma_a^{(p)}[u]=&\lim_{N\to\infty}\frac{1}{N}\ln\overline{ \mathcal{N}_{a} }
    =\frac{1}{2}\log \left(p-1\right) -\frac{p-2}{4p}\eta^{2} + \kappa^{2}\left(1 - \eta^{2}\frac{p-1}{p^{2}}\right)\Theta^{(p)}[u]+ \Phi(\eta) + O\left(\kappa^{3}\right)
\end{split}
\end{equation}
where we have used:
\begin{equation}
    m^{(2)}_{\nabla^{2}U}[\mu_{1,\eta>2}] - \frac{1}{2}\int dx \frac{(\mu^{(2)}_{1,\eta>2})^{2}}{\mu_{0}} -\frac{1}{2} V_{\nabla^{2}U}^{(2)}\left[\mu_{0}\right]  -\frac{1}{4}\lVert\nabla U\left[\mu_{0}\right]\rVert^{2} = 0
\end{equation}
and
\begin{equation}
    \Theta^{(p)}[u] = \frac{1}{2} V_{\nabla^{2}U}^{(p)}\left[\mu_{0}\right]-\frac{1}{2p}\lVert\nabla U\left[\mu_{0}\right]\rVert^{2}.
\end{equation}
where $m_{\nabla^{2}U}^{(p)}$ and $V_{\nabla^{2}U}^{(p)}$ denote the mean and variance of $\nu_{\nabla^{2}U}$ re-scaled by the variance of the random field Hessian, respectively, for the case of a $p$-spin model (see \eqref{eq:Hessian_var_def}).

\subsection{Number of Minima and Maxima for $p=2$} \label{Complexity_index_p_2}
For the case of $p=2$ the expression for the complexity for a given index specializes to:
\begin{equation} \label{eq:complexity_k_result_p_2}
\begin{split}
    \Sigma_a^{(2)}[u]=&
    = \kappa^{2}\left(1 - \frac{\eta^{2}}{4}\right)\Theta^{(2)}[u]+ \Phi(\eta) + O\left(\kappa^{3}\right)
\end{split}
\end{equation}
Note that for $\eta=\pm 2$ ($a=0 , 1$) we have $\Sigma_a^{(2)}=0$. To evaluate the complexity for maxima and minima we should also consider the complexity for $|\eta|>2$ i.e. 
\begin{equation}
    \Sigma_{\mathrm{max}}^{(2)} = \sup_{\eta \ge 2} \kappa^{2}\left(1 - \frac{\eta^{2}}{4}\right)\Theta^{(2)}[u]+ \Phi(\eta)
\end{equation}
Since we know that $\Phi(\eta)$ is non-positive and decreasing monotonically we can deduce that:
\begin{equation}
    \Sigma_{\mathrm{max}}^{(2)} = \Sigma_{\mathrm{min}}^{(2)} = \Sigma_{a=1}^{(2)} = 0
\end{equation}

\newpage

\section{Evaluation of the Mean Number of Critical Points of a Given Index and Energy}
Here we aim to evaluate the number of critical points of a given index $k$ (number of negative eigenvalues of the Hessian). We recall \eqref{eq:fixed_point_k_E_num_def}.
Taking identical steps as for the case of $\overline{ N_{k} }$, we arrive at the average value $\overline{ \left| \det\left( \nabla^{2}H \right) \right|\delta\left( I\left(\nabla^{2} H\right) - k \right)\delta\left(H-E\right)}$. This is evaluated as follows:
\begin{equation}
\begin{split}
     & \overline{ \left| \det\left( \nabla^{2}H \right) \right|  \delta\left( I\left(\nabla^{2} H\right) - k \right)\delta\left(H-E\right)} =  \overline{ \left| \det\left( \nabla^{2} \phi + \kappa\nabla^{2}U\right) \right|\delta\left( I\left(\nabla^{2} \phi + \kappa\nabla^{2}U\right) - k \right) \delta\left(\phi + \kappa U -E\right) } 
     \\
     = & \int_{\mathbb{R}}\frac{\exp{\left(-\frac{Ns^{2}}{2p^2}\right)}}{\sqrt{2\pi p^{2}/N}}\delta\left(-N \frac{s}{p} + \kappa U -E\right)\overline{ \left|\mathrm{det} \tilde{M} + \kappa\nabla^{2}U + s\mathrm{I}_{N-1}\right|\delta\left( I\left(\tilde{M} + \kappa\nabla^{2}U + s\mathrm{I}_{N-1} \right) - k \right) }
\end{split}
\end{equation}
with $\tilde{M}\sim \sqrt{p(p-1)\frac{N-1}{N}}\mathrm{GOE}_{N-1}$ and where $s = - p \, \phi $. Note that and where we have used the conditional distribution of $\nabla^{2}\phi$ given $\phi$ as cited in \ref{item:Random_Hessian_conditional_distribution}. We first evaluate the GOE averaging as in \cite{Arous2021LandscapeManifold} to get:
\begin{align} 
    &\overline{\left|\mathrm{det} \nabla^{2}H\left(\sigma\right)\right|\delta\left( I\left(\nabla^{2} H\right) - k \right) \delta\left(H-E\right)} \\
    &= \int_{\mathbb{R}}\frac{\exp{\left(-\frac{Ns^{2}}{2p^2}\right)}}{\sqrt{2\pi p^{2}/N}}\delta\left(-N \frac{s}{p} + \kappa U -E\right) \delta\left(\int_{-\infty}^{0}\nu_{\nabla^{2}H} \left(x - s\right) dx = k\right) \exp{N \left(\int dx \, \nu_{\nabla^{2}H} \left(x\right) \log\left|x+s\right| +o_N(1)\right)},
\end{align}
where $\nu_{\nabla^{2}H}$ is the widened and shifted semicircular distribution found in \eqref{eq:hessian_spectral_dist}. Note that aside from the delta function constraining the energy, this is the exact same expression as in the case of a fixed index. However, now the integration variable $s$ is constrained twice i.e. unless a specific relation exists between $E$ and $k$ the integral vanishes. This is consistent with the $O(\exp{(-N^{2})})$ concentration between the index and the energy in pure p-spin models \cite{Auffinger2013RandomGlasses}. Thus, we have:
\begin{align*} 
    &\overline{\left|\mathrm{det} \nabla^{2}H\left(\sigma\right)\right|\delta\left( I\left(\nabla^{2} H\right) - k \right) \delta\left(H-E\right)} \\
    &= \delta\left(-N \frac{s^{*}\left(a\right)}{p} + \kappa U -E\right) \exp{N \left(\int dx \, \nu_{\nabla^{2}H} \left(x\right) \log\left|x+s^{*}\left(a\right)\right| - \frac{s^{*}\left(a\right)^{2}}{2 p^{2}}+o_N(1)\right)},
\end{align*}
where, as in the previous section, we have explicitly assumed that $k$ is extensive i.e. the ratio $a = k/N$ is fixed when the large $N$ limit is taken. Also, $s^{*}\left(a\right)$ is determined by the index constraint:
\begin{equation}
    \int_{-\infty}^{0}\nu_{\nabla^{2}H} \left(x - s^{*}\left(a\right)\right) dx = a.
\end{equation}
which is given explicitely by (see derivation in \eqref{eq-eta}):
\begin{equation}
    \eta(a)=-\frac{\kappa\sqrt{p(p-1)} m_{\nabla^{2}U}+s^{*}(a)}{\sqrt{p(p-1)(1+\kappa^2 V_{\nabla^{2}U})}}  \Rightarrow s^{*}(a) = -\eta(a)\sqrt{p(p-1)(1+\kappa^2 V_{\nabla^{2}U})} - \kappa\sqrt{p(p-1)} m_{\nabla^{2}U}
\end{equation}
where the function $\eta(a)$ is defined by:
\begin{equation}
    \int_{-\infty}^{0}\rho_{\mathrm{sc}} \left(x - \eta\left(a\right)\right) dx = a.
\end{equation}
We proceed to the integration over the measure $\mu$ on the sphere:
\begin{align*}
    \overline{ \mathcal{N}_{aN} (E) } = & \frac{C_{N}}{\mathcal{Z}}\int D\left[\mu\right] \delta\left(\int dx \mu\left(x\right) x^{2} - 1\right) \delta\left(-N \frac{s^{*}\left(a\right)}{p} + \kappa N \int dx\, \mu(x) u(x) -E\right)
    \\
    & \times \exp{N\left( \frac{1}{N}\log\overline{\left|\mathrm{det} \nabla^{2}H\left(\sigma\right)\right|\delta\left( I\left(\nabla^{2} H\right) - a N \right)} -\frac{\kappa^{2}}{2 p N} \lVert\nabla U\left[\mu\right]\rVert^{2} - \int dx \mu\log\mu \right)}
\end{align*}
Note that, again, aside from the delta function constraining $a$ in terms of $E$ (for a given position on the $N$-sphere) the result is identical to that obtained for the case of a fixed index. Thus, the optimization  w.r.t $\mu$ (carried out to find the saddle point contribution) would be as in the previous section with an additional Lagrange multiplier, $\gamma$, enforcing the second delta function in the equation above:
\begin{align}
    & \frac{\eta(a)\sqrt{p(p-1)(1+\kappa^2 V_{\nabla^{2}U})} + \kappa\sqrt{p(p-1)} m_{\nabla^{2}U}}{p} + \kappa U -\epsilon = 0
    \\
    & \frac{\eta(a)\sqrt{p(p-1)}\left(1+\kappa^2 \frac{V_{\nabla^{2}U}}{2 }\right) + \kappa\sqrt{p(p-1)} m_{\nabla^{2}U}}{p} + \kappa U -\epsilon = 0
    \\
    & \eta(a)\sqrt{\frac{p-1}{p}} + \eta(a)\kappa^2 \sqrt{\frac{p-1}{p}} \frac{V_{\nabla^{2}U}}{2} + \kappa\sqrt{\frac{p-1}{p}}m_{\nabla^{2}U} + \kappa U -\epsilon = 0
    \\
    & \kappa \sqrt{\frac{p-1}{p}} \left( \frac{1}{\kappa}\left(\eta(a) - \sqrt{\frac{p}{(p-1)}}\epsilon\right) + \eta(a)\kappa \frac{V_{\nabla^{2}U}}{2 } + m_{\nabla^{2}U} + \sqrt{\frac{p }{p-1}}U\right) = 0
\end{align}
That is, the functional for optimization is:
\begin{equation}
    \begin{split}
        S_{a,E}[\mu] = &  \beta \left(\int dx \mu\left(x\right) x^{2} - 1\right) + \gamma\left(\frac{1}{\kappa}\left(\eta(a) - \sqrt{\frac{p}{(p-1)}}\epsilon\right) + \frac{\kappa \eta(a)}{2 }  V_{\nabla^{2}U}[\mu] + m_{\nabla^{2}U}[\mu] + \sqrt{\frac{p }{p-1}}U[\mu]  \right) 
        \\
        & +\frac{1}{N}\log\overline{\left|\mathrm{det} \nabla^{2}H\left(\sigma\right)\right|\delta\left( I\left(\nabla^{2} H\right) - a N \right)} -\frac{\kappa^{2}}{2 p N} \lVert\nabla U\left[\mu\right]\rVert^{2} - \int dx \mu\log\mu +\alpha \left(\int dx \mu\left(x\right) - 1\right) 
    \end{split}
\end{equation}
Where $\epsilon = E/N$ 
Taking the functional derivative w.r.t $\mu$ of the functional above, substitution of explicit expressions for the the disorder average of the determinant in \eqref{eq:log_det_k} and appropriate perturbative expansions \eqref{eq:perturbative_expansion_def} we get the following equations for $\mu$:
\begin{equation}
    \begin{split}
        & O(\kappa^{0}):\, \alpha_{0}+\beta_{0}x^2-\log (\mu_{0}(x)) = 0
        \\
        & O(\kappa^{1}):\, \alpha_{1}+\beta_{1}x^2-\frac{\mu_{1}(x)}{\mu_{0}(x)} - \sqrt{\frac{p-1}{p^{3}}}\eta(a)\left(u''(x)-x u'(x)\right) + \gamma_{1}\left(\frac{u''(x)-x u'(x)}{\sqrt{p(p-1)}} + \sqrt{\frac{p }{p-1}} u(x)\right)=0.
    \end{split} 
\end{equation} 
Where we have implicitly assumed that $\int \mu_{0}(x) u(x) dx = \int \mu_{0}(x) u''(x) dx =  0$ and that we are interested only in index and energy values very close to the deterministic relation between energy and index in the unperturbed case - up to $O(\kappa^{2})$:
\begin{equation}
    \eta(a) - \sqrt{\frac{p}{p-1}}\epsilon = \kappa^{2}\Delta 
\end{equation}
with $\Delta = O(1)$. We note here that requiring an $O(\kappa)$ shift would give rise to an $O(1)$ change in $\mu_{0}$ and thus would result in $O(1)$ negative complexity. This calculation is outside the scope of this work. Moreover, for $\Delta = O(1)$ there are exactly zero critical points. 

Solving these equations, along with the constraints as above, gives the following solutions:
\begin{equation}
    \begin{split}
        \mu_{0}(x) = & \frac{1}{\sqrt{2\pi}}e^{-x^{2}/2}
        \\
        \mu_{1}(x) = & \mu_{0}(x)\,\left[-\sqrt{\frac{p-1}{p^{3}}}\eta \left( u''(x)-x u'(x)\right)   + \gamma_{1} \left( \frac{u''(x)-x u'(x)}{\sqrt{p(p-1)}} +\sqrt{\frac{p}{p-1}}u(x)\right)\right]
    \end{split}
\end{equation}
Note that the first term in $\mu_{1}$ is identical to the $\mu_{1}$ at the saddle point for the case of fixed index (see \eqref{eq:mu_k}). We denote this decomposition by
\begin{align}
     & \mu_{1,k,E}(x) =  \mu_{1,k}(x) + \gamma_{1}\tilde{\mu}(x)
     \\
     & \mu_{1,k}(x) = -\mu_{0}(x) \sqrt{\frac{p-1}{p^{3}}}\eta \left(  u''(x)-x u'(x)\right)
     \\
     & \Tilde{\mu}(x) = \mu_{0}(x) \left( \frac{u''(x)-x u'(x)}{\sqrt{p(p-1)}} +\sqrt{\frac{p}{p-1}} u(x)\right)
\end{align}
The expression above is consistent with the normalisation and second moment constraints. To find $\gamma_{1}$ we impose the energy constraint (using $U[\mu_0]=0$ and $m[\mu_0]=0$):
\begin{align}
    & \kappa \Delta  + \frac{\kappa \eta(a)}{2 }  V_{\nabla^{2}U}[\mu_{0}] + \kappa m_{\nabla^{2}U}[\mu_{1}] + \kappa \sqrt{\frac{p }{p-1}}U[\mu_{1}] = 0
    \\
    & \Delta + \frac{ \eta(a)}{2 }  \frac{1}{p(p-1)}\overline{u''^{2}} + m_{\nabla^{2}U}[\mu_{1,k}] + \gamma_{1}m_{\nabla^{2}U}[\Tilde{\mu}] + \sqrt{\frac{p }{p-1}}U[\mu_{1,k}] + \gamma_{1}\sqrt{\frac{p }{p-1}}U[\Tilde{\mu}] = 0
    \\
    & \gamma_{1} = -\frac{\Delta + \frac{ \eta(a)}{2 }  \frac{1}{p(p-1)}\overline{u''^{2}} + m_{\nabla^{2}U}[\mu_{1,k}]  + \sqrt{\frac{p }{p-1}}U[\mu_{1,k}]}{\sqrt{\frac{p }{p-1}}U[\Tilde{\mu}] + m_{\nabla^{2}U}[\Tilde{\mu}]}
\end{align}
where we use the over-bar notation over functions of coordinates i.e. $\overline{f(x)}$ to denote weighed averaging w.r.t the standard Gaussian measure $\mu_0$: $\overline{f(x)} = \int f(X) \mu_0(x) dx$. Let us now evaluate the functionals $m$ and $U$: 
\begin{align}
     m_{\nabla^{2}U}[\mu_{1,k}] &= -\sqrt{\frac{p-1}{p^{3}}}\eta \frac{1}{\sqrt{p(p-1)}} \overline{\left(  u''(x)-x u'(x)\right)^{2}} = \frac{\eta}{p^{2}}\overline{u''^{2}-2 x u'' u' +x^{2}u'^{2}} = \frac{\eta}{p^{2}}\overline{u''^{2}-2 x u'' u' + u'^{2}+ 2 x u' u''}\nonumber
    \\
    &= -\frac{\eta}{p^{2}}\left(\overline{u''^{2}}+\overline{u'^{2}}\right)
    \\
     \sqrt{\frac{p }{p-1}}U[\mu_{1,k}] &= -\eta\sqrt{\frac{p }{p-1}} \sqrt{\frac{p-1}{p^{3}}} \overline{u(x)\left(  u''(x)-x u'(x)\right)} = \frac{\eta}{p}\overline{u u'' - u'^{2} - u u''} = \frac{\eta}{p}\overline{u'^{2}}
\end{align}
where in the last equality we have used $\overline{x^{2}u'^{2}} = \overline{(x u'^{2})'} = \overline{u'^{2} + 2 x u' u''}$, 
where we have used $V[\mu_{0}] = \frac{1}{p(p-1)}\overline{u''^{2}}$ since we can fix $\overline{u''}=0$ (by adding a quadratic term to the potential).
Gathering the terms we get the expression for $\gamma_{1}$:
\begin{equation}
    \gamma_{1} = \frac{\frac{1}{\kappa^{2}}\left(\eta(a) - \sqrt{\frac{p}{p-1}}\epsilon\right) + \eta(a)\left(\left(\frac{1}{2 p(p-1)} - \frac{1}{p^{2}}\right) \overline{u''^{2}} + \left(-\frac{1}{p^{2}} + \frac{1}{p}\right)\overline{u'^{2}}\right)}{\sqrt{\frac{p }{p-1}}U[\Tilde{\mu}] + m_{\nabla^{2}U}[\Tilde{\mu}]}
\end{equation}


Now we can substitute $\gamma_1$ substitute it in the expression for the number of critical points (see \eqref{eq:num_points_k_pert_expansion} to get:
\begin{equation}
    \Sigma_a^{(p)}(E)=\lim_{N\to\infty}\frac{1}{N}\ln\overline{ \mathcal{N}_{a}(E) } = \Sigma_a^{(p)} + \kappa^{2}\left(\frac{p-1}{p}\eta m_{\nabla^{2}U}^{(p)}[\gamma_{1} \tilde{\mu}] - \frac{1}{2}\left(\gamma_{1}^2\int \frac{\tilde{\mu}^{2}}{\mu_{0}} dx + 2\gamma_{1}\int \frac{\tilde{\mu} \, \mu_{1,k}}{\mu_{0}} dx \right)\right) + O(\kappa^{3})
\end{equation}
where $\Sigma_a^{(p)}$ is the complexity for index $k=a N$ given in \eqref{eq:complexity_k_result}. We can show that:
\begin{align}
    & \frac{p-1}{p}\eta m_{\nabla^{2}U}^{(p)}[\tilde{\mu}] = \int \frac{\tilde{\mu} \, \mu_{1,k}}{\mu_{0}} dx
    \\
    & \int \frac{\tilde{\mu}^{2}}{\mu_{0}} dx = \sqrt{\frac{p }{p-1}}U[\Tilde{\mu}] + m_{\nabla^{2}U}[\Tilde{\mu}] = \overline{\left( \frac{u''(x)-x u'(x)}{\sqrt{p(p-1)}} +\sqrt{\frac{p}{p-1}} u(x)\right)^{2}} = \frac{\overline{u''^{2}}+\overline{u'^{2}}}{p(p-1)} + \frac{p}{p-1}\overline{u^{2}} - 2\frac{\overline{u'^{2}}}{p-1}
\end{align}
which leads to:
\begin{equation}
    \Sigma_a^{(p)}(E) = \Sigma_a^{(p)} - \frac{\kappa^{2}\gamma_{1}^{2}}{2}\int \frac{\tilde{\mu}^{2}}{\mu_{0}} dx + O(\kappa^3) 
\end{equation}
Further substitution of the expression for $\gamma_{1}$ gives:
\begin{equation}
    \Sigma_a^{(p)}(\epsilon) = \Sigma_a^{(p)} - \frac{\kappa^{2}}{2}\left(\frac{1}{\kappa^{2}}\left(\eta(a) - \epsilon\sqrt{\frac{p}{p-1}}\right) + \eta(a)\tilde{\Delta} \right)^{2} + O(\kappa^{3})
\end{equation}
with $\epsilon = E/N$ and where $\Tilde{\Delta}$ is given by:
\begin{equation} \label{eq:Delta_tilde_def}
    \Tilde{\Delta} = \left(\frac{1}{2 p(p-1)} - \frac{1}{p^{2}}\right) \overline{u''^{2}} + \left(-\frac{1}{p^{2}} + \frac{1}{p}\right)\overline{u'^{2}}
\end{equation}
Note that this analysis only considered critical points with an index that diverges with $N$, thus the maximal (or minimal) energies where this complexity is nonzero cannot be identified as the global maxima (minima) values. This is especially true for $p>2$ where it is a known result the global maxima (minima) have a hessian spectrum which is gapped from zero \cite{Auffinger2013RandomGlasses}.

\appendix

\renewcommand{\theequation}{A\arabic{equation}}
\section{Distribution of Random Fields} \label{Appendix A: Dist of Random Fields}
In \cite{Auffinger2013RandomGlasses} Lemma 1, the covariances of the field $\phi$, and its covariant derivatives $\nabla \phi$, and $\nabla^{2}\phi$ at a given point were derived for the of $p$-spin spherical models and are given by:
\begin{equation}
    \begin{split}
        \mathbb{E}\left[\phi^{2}\right] & = N 
        \\
        \mathbb{E}\left[\left(\nabla \phi\right)_{m} \phi\right] & = \mathbb{E}\left[\left(\nabla \phi\right)_{m} \left(\nabla^{2} \phi\right)_{np}\right] = 0
        \\
        \mathbb{E}\left[\left(\nabla \phi\right)_{m}\left(\nabla \phi\right)_{n}\right] & =  p  \delta_{mn}
        \\
        \mathbb{E}\left[\phi \left(\nabla^{2} \phi\right)_{mn}\right] & = -p  \delta_{mn}
        \\
        \mathbb{E}\left[\left(\nabla^{2} \phi\right)_{mn} \left(\nabla^{2} \phi\right)_{lp}\right] & = \frac{p\left(p-1\right)}{N }\left(\delta_{ml}\delta_{np} + \delta_{mp}\delta_{nl}\right) + \frac{p^{2}}{N }\delta_{mn}\delta_{lp}
    \end{split}
\end{equation}
Note that here $m=1,\ldots,N-1$ and the coordinates of the gradient and the Hessian are with respect to an orthonormal system in the tangent space to the sphere at $p$.

\renewcommand{\theequation}{B\arabic{equation}}
\section{$N$-Sphere covariant Hessian for Single Particle Potential} \label{Appendix B N-Sphere Covariant Hessian Single Particae}
The covariant gradient on the sphere can be written concisely as a Euclidean gradient followed by a projection to the sphere tangent surface:
\begin{equation}
    \nabla_{\mathbb{S}^{N}}  = \left(I_{N}-\hat{\sigma}\otimes\hat{\sigma}\right) \cdot \nabla_{\mathbb{R}^{N}}
\end{equation}
where we have denoted for clarity $\nabla_{\mathbb{S}^{N}}$ as the covariant gradient on the sphere and $\nabla_{\mathbb{R}^{N}}$ the Euclidean gradient in $\mathbb{R}^{N}$. Also, $I_{N}$ denotes the unit operator in $\mathbb{R}^{N}$ and $\hat{\sigma}$ is the radial unit vector. The covariant Hessian is a result of  two repeated applications of the covariant Hessian:
\begin{equation}
    \nabla_{\mathbb{S}^{N}}^{2}U = \left(I_{N}-\hat{\sigma}\otimes\hat{\sigma}\right) \cdot \nabla_{\mathbb{R}^{N}} \left(\left(I_{N}-\hat{\sigma}\otimes\hat{\sigma}\right) \cdot \nabla_{\mathbb{R}^{N}} U \right)
\end{equation}
expanding this expression we get
\begin{equation}
\label{eq-B3}
    \begin{split}
        \nabla_{\mathbb{S}^{N}}^{2}U = \left(I_{N}-\hat{\sigma}\otimes\hat{\sigma}\right)\cdot \left(\nabla_{\mathbb{R}^{N}}^{2}U -\nabla_{\mathbb{R}^{N}}\left(\hat{\sigma}\otimes\hat{\sigma}\cdot \nabla_{\mathbb{R}^{N}} U \right)\right).
    \end{split}
\end{equation}
We now turn to evaluate the last term on the right hand side in \eqref{eq-B3}  for the case of single particle $U$:
\begin{equation}
\begin{split}
    & \left(\nabla_{\mathbb{R}^{N}}\left(\hat{\sigma}\otimes\hat{\sigma}\cdot \nabla_{\mathbb{R}^{N}} U \right)\right)_{ij} =  \frac{\partial}{\partial x_{i}} \sum_{m} \frac{x_{j}x_{m}}{\sum_{n}x_{n}^{2}} \frac{\partial U}{\partial x_{m}} = 
    \\
    & \sum_{m} \left( \frac{\delta_{ij}x_{m}+x_{j}\delta_{im}}{\sum_{n}x_{n}^{2}} - 2\frac{x_{i}x_{j}x_{m}}{\left(\sum_{n}x_{n}^{2}\right)^{2}} \right) \frac{\partial U}{\partial x_{m}} + \sum_{m} \frac{x_{j}x_{m}}{\sum_{n}x_{n}^{2}} \frac{\partial^{2} U}{\partial x_{m} \partial x_{i}} = 
    \\
    & \frac{1}{\sqrt{\sum_{n}x_{n}^{2}}}\left(\delta_{ij}\sum_{m}\hat{\sigma}_{m} \frac{\partial U}{\partial x_{m}} + \hat{\sigma}_{j}\frac{\partial U}{\partial x_{i}} - 2\left(\hat{\sigma}\otimes\hat{\sigma}\right)_{ij} \sum_{m} \hat{\sigma}_{m}\frac{\partial U}{\partial x_{m}} \right) + \sum_{m} \left(\hat{\sigma}\otimes\hat{\sigma}\right)_{jm} \frac{\partial^{2} U}{\partial x_{m} \partial x_{i}}
\end{split}
\end{equation}
where we have used the following identity:
\begin{equation}
    \frac{\partial}{\partial x_{i}} \left(\frac{x_{j}x_{m}}{\sum_{n}x_{n}^{2}}\right) = \frac{\delta_{ij}x_{m}+x_{j}\delta_{im}}{\sum_{n}x_{n}^{2}} - 2\frac{x_{i}x_{j}x_{m}}{\left(\sum_{n}x_{n}^{2}\right)^{2}}.
\end{equation}
Now, we can write this expression in vector notation as follows:
\begin{equation}
    \nabla_{\mathbb{R}^{N}}\left(\hat{\sigma}\otimes\hat{\sigma}\cdot \nabla_{\mathbb{R}^{N}} U \right) = \frac{1}{\sqrt{N}}\left(\left(\hat{\sigma}\cdot\nabla_{\mathbb{R}^{N}} U\right)I_{N} + \hat{\sigma}\otimes\nabla_{\mathbb{R}^{N}} U - 2 \left(\hat{\sigma}\cdot\nabla_{\mathbb{R}^{N}} U\right)\hat{\sigma}\otimes\hat{\sigma}\right) + \left(\hat{\sigma}\otimes\hat{\sigma}\right) \cdot \nabla_{\mathbb{R}^{N}}^{2}U
\end{equation}
where we have used the fact that the derivative is calculated on $\sqrt{N}\mathbb{S}^{N-1}$ i.e $\sum_{n} x_{n}^{2} = N$. For the purpose of the asymptotic calculation carried out in this work all terms of rank $O\left(1\right)$ can be neglected. Observation of the above expressions yields that:
\begin{equation}
    \nabla_{\mathbb{S}^{N}}^{2}U = \nabla_{\mathbb{R}^{N}}^{2}U - \frac{1}{\sqrt{N}}\left(\hat{\sigma}\cdot\nabla_{\mathbb{R}^{N}} U\right)I_{N} + M_{1}
\end{equation}
where $M_{1}$ denotes a matrix of rank $O\left(1\right)$.

\renewcommand{\theequation}{C\arabic{equation}}
\section{Evaluating The Density of the Empirical Spectral Distribution of $\nabla^{2}H = \nabla^{2}\phi + \kappa\nabla^{2}U$} \label{Appendix C free convolution}

Denoting by $G\left(z\right)$ the Stieltjes transform of $\rho_{\mathrm{sc}} \boxplus \nu_{\kappa\nabla^{2}U}$, $G$ is known to obey the following implicit formula \cite{Biane1997OnDistribution} Proposition 2:
\begin{equation} \label{eq:free_convolution_transform_implicit_eq}
    G\left(z\right) = \int \frac{\nu_{\kappa\nabla^{2}U}\left(x\right) dx}{z  - t^{2}G\left(z\right)-x} = \int \frac{\frac{1}{\kappa}\nu_{\nabla^{2}U}\left(\frac{x}{\kappa}\right) dx}{z - t^{2}G\left(z\right)-x}, \quad z\in \mathbf{C}_+
\end{equation}
where $t^{2} = p\left(p-1\right)$ and we have used the identity $\nu_{\kappa\nabla^{2}U} = \frac{1}{\kappa}\nu_{\nabla^{2}U}\left(\frac{x}{\kappa}\right)$. To retrieve  $\nu_{\nabla^{2}H} = \rho_{\mathrm{sc}} \boxplus \nu_{\kappa\nabla^{2}U}$ one employs the inverse transform:
\begin{equation}
    \psi \left(x\right) = \frac{1}{\pi} 
    \lim_{\epsilon \rightarrow 0^{+}} \mathrm{Im}\, G\left(x + \mathrm{i}\epsilon\right).
\end{equation}

\subsubsection{Evaluating the Free Convolution with a Small Perturbation}
Here, we employ perturbation theory in $\kappa$ to solve the implicit equation \eqref{eq:free_convolution_transform_implicit_eq}. First we make the change of variables $s = {x}/{\kappa}$ followed 
by the change of variables $s = m_{\nabla^{2}U} + s'$ with $m_{\nabla^{2}U}$ denoting the expectation value of $\nu_{\nabla^{2}U}$,
\begin{equation}
    \int \frac{\frac{1}{\kappa}\nu_{\nabla^{2}U}\left(\frac{x}{\kappa}\right) dx}{z - t^{2}G\left(z\right)-x} = \int \frac{\nu_{\nabla^{2}U}\left(s\right) ds}{z  - t^{2}G\left(z\right)-\kappa s} = \int \frac{\nu_{\nabla^{2}U}\left( m_{\nabla^{2}U}+s'\right) ds'}{z  - t^{2}G\left(z\right)-\kappa\left( m_{\nabla^{2}U} + s'\right) }.
\end{equation}
Now  we can expand the integral in a power series in $\kappa$ as follows:
\begin{equation}
\begin{split}
    & \int \frac{\nu_{\nabla^{2}U}\left(s' +  m_{\nabla^{2}U}\right) ds'}{z  - t^{2}G\left(z\right)-\kappa\left( m_{\nabla^{2}U} + s'\right)} =  \frac{1}{z  - t^{2}G(z) - \kappa m_{\nabla^{2}U}}\int \frac{\nu_{\nabla^{2}U}\left(s' +  m_{\nabla^{2}U}\right) ds'}{1-\kappa\frac{s'}{z  - t^{2}G-\kappa m_{\nabla^{2}U}}} 
    \\
    = & \frac{1}{z  - t^{2}G-\kappa m_{\nabla^{2}U}} + \frac{\kappa^{2}}{\left(z  - t^{2}G-\kappa m_{\nabla^{2}U}\right)^{3}}\int s'^{2}\nu_{\nabla^{2}U}\left(s' +  m_{\nabla^{2}U}\right)ds' + O\left(\kappa^{3}\right)
\end{split}
\end{equation}
where we have used the simple fact that $\int s'\nu_{\nabla^{2}U}\left(s' +  m_{\nabla^{2}U}\right)ds'=0$.  Reordering  \eqref{eq:free_convolution_transform_implicit_eq} results in:
\begin{equation} \label{eq:free_conv_pert_expansion}
    \left(z-\kappa m_{\nabla^{2}U}\right)G - t^{2}G^{2} - 1 = \frac{\kappa^{2}}{\left(z -\kappa m_{\nabla^{2}U} - t^{2}G\right)^{2}}\int s'^{2}\nu_{\nabla^{2}U}\left(s' +  m_{\nabla^{2}U}\right) ds' + O\left(\kappa^{3}\right).
\end{equation}
We solve this perturbative equation, up to order $\kappa^{2}$, by introducing the following ansatz:
\begin{equation}
     G\left(z\right) =  \frac{z  - \kappa m_{\nabla^{2}U}+ \sqrt{\left(z-\kappa m_{\nabla^{2}U}\right)^{2} - 4T^{2}}}{2T^{2}}
\end{equation}
where we expand the support $T$ in a power series:
\begin{equation}
    T^{2} = t^{2} + \kappa T_{1} + \kappa^{2}T_{2}+O\left(\kappa^{3}\right).
\end{equation}
Substitution in the equation above results in:
\begin{equation}
    \begin{split}
        T_{1} & = 0
        \\
        T_{2} & = \int s'^{2}\nu_{\nabla^{2}U}\left(s' +  m_{\nabla^{2}U}\right) ds' \equiv V_{\nabla^{2}U}.
    \end{split}
\end{equation}
It is seen that up to second order in $\kappa$ the free convolution is a slightly wider semicircle distribution with a width of $t^{2} + \kappa^{2}V_{\nabla^{2}U}$:
\begin{equation}
    \nu_{\nabla^{2}H}\left(x\right) = 1_{x\in\left[-2T+\kappa m_{\nabla^{2}U},2T+\kappa m_{\nabla^{2}U}\right]}\frac{1}{2\pi\left(t^{2} + \kappa^{2}V_{\nabla^{2}U}\right)}\sqrt{-\left(x-\kappa m_{\nabla^{2}U}\right)^{2} + 4\left(t^{2} + \kappa^{2}V_{\nabla^{2}U}\right)}.
\end{equation}
\section{Positivity of $\Theta^{(2)}\left[ f(x) \right]$ for potentials with $f'(0) = 0$} \label{Appendix E positivity of theta}
\renewcommand{\theequation}{D\arabic{equation}}

Returning to our main model (i.e. $p=2$), let us consider the physical interpretation of the system's state in each of the branches. We start by identifying the families of potentials which give rise to each of the branches. We consider a general case of a confining potential given by a power series, convergent for all $x\in \mathbb{R}$:
\begin{equation}
    u\left(x\right) = \sum_{n=1}^{\infty} a_{n} x^{n}
\end{equation}
$\Theta \left[u\right]$ for this power series is given by:
\begin{eqnarray} \label{eq:power_series_Theta}
     \Theta \left[\sum_{n=1}^{\infty} a_{n} x^{n}\right]  &=& - \frac{a_{1}^2}{4} - a_{1}\sum_{n \ge 3} A_n a_n + \sum_{n \ge 3}B_n a_n^2 \nonumber \\
    & &  
 + \sum_{m \ge 3}\sum_{n \ge 3} D_{n,m} a_m a_n
\end{eqnarray}
where $A_n,B_n,D_{m,n}>0$, and their expressions are given below. Note that, as expected, the term $a_{2}$ does not affect the complexity since it corresponds to a constant function on the sphere. We draw the following conclusions from this expression: (i) if $a_1=0$, then $\Theta \left[u\right]$ is positive (see proof in SI) and thus the total complexity is positive as well (see \ref{eq:complexity_main_result}). (ii) If the potential is linear (only $a_1$ is nonzero) then $\Theta \left[u\right]$ is negative and the total complexity is zero to leading order. 

Here we show that $\Theta^{(2)}[u(x)]$ is positive when $u'(0)=0$ or equivalently $a_{1}=0$ in the Taylor expansion of $u(x)$:
\begin{equation}
    u\left(x\right) = \sum_{n=1}^{\infty} a_{n} x^{n}
\end{equation}
Explicit integration gives
\begin{equation}
     \Theta \left[\sum_{n=1}^{\infty} a_{n} x^{n}\right]  = - \frac{a_{1}^2}{4} - a_{1}\sum_{n \ge 3} A_n a_n + \sum_{n \ge 3}B_n a_n^2 + \sum_{m \ge 3}\sum_{n \ge 3} D_{n,m} a_m a_n
\end{equation}
where we have denoted:
\begin{align}
    A_{n} = & \frac{1}{4\sqrt{2 \pi }} 2^{n/2} c_{n+1} n \Gamma \left(\frac{n}{2}\right)
    \\
    B_{n} = & \frac{1}{4\sqrt{2 \pi }}2^{n-\frac{3}{2}} (n-2)^2 n^2  \Gamma \left(n-\frac{3}{2}\right)
    \\
    D_{n,m} = & \frac{1}{4\sqrt{2 \pi }}(m-2) m (n-2) n 2^{\frac{1}{2} (m+n-3)} c_{n+m} \Gamma \left(\frac{1}{2} (m+n-3)\right)
\end{align}
with $c_{n} = \left((-1)^{n}+1\right)$. 

Next, we show that $\Theta[u]>0$ for $a_{1}=0$. As the term $\sum_{n \ge 3}B_n a_n^2$ is obviously positive it is left to show that $\sum_{m \ge 3}\sum_{n \ge 3} D_{n,m} a_m a_n > 0$. Using the inequality $\Gamma \left(\frac{1}{2} (m+n-3)\right) > \Gamma \left(\frac{1}{2} (m-3/2)\right) \Gamma \left(\frac{1}{2} (n-3/2)\right)$ we can write:
\begin{equation}
\begin{split}
    & \sum_{m \ge 3}\sum_{n \ge 3} D_{n,m} a_m a_n = \sum_{m \ge 3}\sum_{n \ge 3}m (m-2) n (n-2)   2^{\frac{1}{2} (m+n-3)} c_{n+m} \Gamma \left(\frac{1}{2} (m+n-3)\right) a_m a_n >
    \\
    & \left(\sum_{n \ge 3}a_n n (n-2)   2^{\frac{1}{2} (n-3/2)} \Gamma \left(\frac{1}{2} (n-3/2)\right)\right)^{2}\\
    &+ \left(\sum_{n \ge 3}a_n n (n-2) (-1)^{n} 2^{\frac{1}{2} (n-3/2)} \Gamma \left(\frac{1}{2} (n-3/2)\right)\right)^{2}
\end{split}
\end{equation}
and we have the desired result. 
